\documentclass[preprint,12pt]{elsarticle}
\usepackage{amsmath}
\usepackage{graphicx,fancyhdr}
\usepackage{graphicx,amsmath,bm}
\usepackage{amssymb,amscd}
\usepackage{caption2,color}
\usepackage{amssymb}

\journal{}

\begin{document}

\begin{frontmatter}

\title{\Large\bf Monotonic and  nonmonotonic immune responses in viral
infection systems \tnoteref{t1}}
 \tnotetext[t1]{This work is supported by NSFC (No. U1604180), Key Scientific and Technological Research Projects in
Henan Province (No.192102310089), Foundation of Henan Educational
Committee (No.19A110009) and Grant of Bioinformatics Center of Henan
University (No.2018YLJC03).}
\author[els]{Shaoli Wang \corref{cor1}}
 \ead{wslheda@163.com }
\author[els]{Huixia Li}
\author[wlu]{Fei Xu}
\ead{fxu.feixu@gmail.com}

 \cortext[cor1]{Corresponding author.}

\address[els]{School of Mathematics and Statistics, Bioinformatics Center,
Henan University, Kaifeng 475001, Henan, PR China}
\address[wlu]{   Department of Mathematics, Wilfrid Laurier University,
Waterloo, Ontario, N2L 3C5, Canada}

\begin{abstract}

In this paper, we study two-dimensional, three-dimensional monotonic
and nonmonotonic immune responses in viral infection systems. Our
results show that the viral infection systems with monotonic immune
response has no bistability appear. However, the systems with
nonmonotonic immune response has bistability appear under some
conditions. For immune intensity, we got two important thresholds,
post-treatment control threshold and elite control threshold. When
immune intensity is less than post-treatment control threshold, the
virus will be rebound. The virus will be under control when immune
intensity is larger than elite control threshold. While between the
two thresholds is a bistable interval. When immune intensity is in
the bistable interval, the system can have bistability appear.
Select the rate of immune cells stimulated by the viruses as a
bifurcation parameter for nonmonotonic immune responses, we prove
the system exhibits saddle-node bifurcation and transcritical
bifurcation.

\end{abstract}

\begin{keyword}
Monotonic immune response; Nonmonotonic immune response;
Post-treatment control threshold;
 Elite control threshold; Bistability; Saddle-node bifurcation; Transcritical bifurcation

\MSC 35B35 \sep 35B40 \sep 92D25
\end{keyword}

\end{frontmatter}

\section{Introduction}
During the process of viral infection, the host is induced which is
initially rapid and nonspecific (natural killer cells, macrophage
cells, etc.) and then delayed and specific (cytotoxic T lymphocyte
cells, antibody cell). But in most virus infections, cytotoxic T
lymphocyte (CTL) cells which attack infected cells and antibody
cells which attack viruses, play a critical part in antiviral
defense. Some researchers have studied some models about virus
dynamics within-host and immune response, \cite{[1],[2],[3],[4],[5]}
and others don't contain the immune responses.
\cite{[6],[7],[8],[9],[10],[11]}

In order to investigate the role of the population dynamics of viral
infection with CTL response, Nowak and Bangham (see e.g. Refs
\cite{[12]}) constructed a mathematical model describing the basic
dynamics of the interaction between activated CD4$^{+}$ T cells,
$x(t)$, infected CD4$^{+}$ T cells, $y(t)$, viruses, $v(t)$ and
immune cells, $z(t)$.
  $$\left\{\begin
 {array}{l l}
\frac{dx}{dt}=s-dx-(1-\epsilon)\beta xy, \\
\frac{dy}{dt}=(1-\epsilon)\beta xy-ay-pyz,\\
\frac{dz}{dt}=f(y)z-bz,
  \end {array}
  \right. \eqno(1.1) $$
where $f(y)$ is a continuously differentiable function defined on
$[0,
  +\infty)$ and satisfies
  $$f^{'}(y)>0,~ f(0)=0 ~\mbox{and} ~f(y)\leq My ~\mbox{for some positive parameter } M. \eqno(1.2) $$

For example, $f(y)=cy$ or $f(y)=\frac{cy}{1+\alpha y}$ is the common
monotonic immune response in viral infection systems.
\cite{[15],[20]} In 1968, Andrews (see e.g. Refs \cite{[13]})
suggested Monod-Haldane function
$$f(y)=\frac{cy}{\alpha+\gamma y+y^2},$$
then, Sokol and Howell (see e.g. Refs \cite{[14]}) proposed a
simplified Monod-Haldane function
$$f(y)=\frac{cy}{\alpha+y^2},$$
as nonmonotonic functions in chemostat systems. The nonmonotonic
functions are also discussed in predator-prey system.
\cite{[16],[17],[18]} Wang et al (see e.g. Refs \cite{[26]})
proposed oxidative stress in a HIV infection model and the immune
function is a Monod-Haldane function. Thus we chose
$\frac{cyz}{\alpha+\gamma y+y^2}$  as the nonmonotonic immune
response in the following system.
$$\left\{\begin
 {array}{l l}
\frac{dx}{dt}=s-dx-(1-\epsilon)\beta xy=g_1, \\
\frac{dy}{dt}=(1-\epsilon)\beta xy-ay-pyz=g_2,\\
\frac{dz}{dt}=\frac{cyz}{\alpha+\gamma y+y^2}-bz=g_3.
  \end {array}
  \right. \eqno(1.3) $$
Activated CD4$^{+}$ T cells are generated at a rate $s$, die at a
rate $d$, and become infected CD4$^{+}$ T cells at a rate
$(1-\epsilon)\beta xy$. Infected CD4$^{+}$ T cells die at a rate $a$
and are killed by immune cells at a rate $pyz$.
$\frac{cyz}{\alpha+\gamma y+y^2}$ represents the immune cells
stimulated by the viruses and die at a rate $b$. All the parameters
are positive.

The rest of this paper is organized as follows. The viral infection
system with monotonic immune response is carried out in section 2.
The stability analysis, bifurcation analysis and numerical
simulations of nonmonotonic immune response is carried out in
Section 3. In section 4, we analyze the 2D-viral infection system
with monotonic immune response. In section 5, we analyze the
stability and bifurcation of the 2D-viral infection system with
monotonic immune response and carry out numerical simulations. In
section 6, we conclude the paper with discussions.

\section{Viral infection system with monotonic immune response}
System (1.1) always has an uninfected steady equilibrium
$E^{(1)}_{0}=(x^{(1)}_{0},0,0)$, and if $\mathcal
{R}^{(1)}_0>1>\mathcal {R}^{(1)}_{*}$, system (1.1) also has an
immune-free equilibrium $E^{(1)}_{1}=(x^{(1)}_{1}, y^{(1)}_{1}, 0)$;
If $\mathcal {R}^{(1)}_0>\mathcal {R}^{(1)}_{*}>1$ system (1.1) has
three equilibria $E^{(1)}_{0}$, $E^{(1)}_{1}$ and
$E^{(1)}_*=(x^{(1)}_*, y^{(1)}_*, z^{(1)}_*)$, where
$$\begin
 {array}{ll}
 x^{(1)}_{0}=\frac{s}{d},\\\vspace{2\jot}
 x^{(1)}_{1}=\frac{a}{\beta(1-\epsilon)},\\\vspace{2\jot}
  y^{(1)}_{1}=\frac{d(R^{(1)}_{0}-1)}{\beta(1-\epsilon)},\\\vspace{2\jot}
x^{(1)}_*=\frac{s}{ d+(1-\epsilon)\beta y^{(1)}_*},\\\vspace{2\jot}
\displaystyle y^{(1)}_*=f^{-1}(b),\\\vspace{2\jot} \displaystyle
z^{(1)}_*=\frac{a(R^{(1)}_{*}-1)}{p}.\end {array}
$$

The basic reproductive number is given as
$$\mathcal
{R}^{(1)}_0=(1-\epsilon)\beta \frac{s}{d}\frac{1}{a} 1=\frac{s\beta
(1-\epsilon)}{ad}.$$ Because $(1-\epsilon)\beta
\frac{s}{d}\frac{1}{a}$ is the basic reproductive number of the
model with the bilinear incidence $\beta xy$, ${R}^{(1)}_0$ gives
the basic reproductive number of system (1.1) with the constant
function response.

The basic immune reproductive number is $$\mathcal
{R}^{(1)}_{*}=\frac{s\beta (1-\epsilon)}{ad+a\beta
(1-\epsilon)y^{(1)}_*}.$$ This ratio describes the average number of
newly infected cells generated form on infected cells at the
beginning of the infectious process.

Let $\tilde{E}$ be any arbitrary equilibrium of system (1.1). The
Jacobian matrix associated with the system is
$$J_{1}=\left[
\begin{array}{cccc}
 -d-\beta(1-\epsilon)y  &  -\beta(1-\epsilon)x  & 0 \\
\beta(1-\epsilon)y   & \beta(1-\epsilon)x-a-pz &-py  \\
0   & f^{'}(y)z & f(y)-b \\
\end{array}
\right].$$ The characteristic equation of the linearized system of
(1.1) at $\tilde{E}$ is given by $\left|\lambda I-J_{1} \right|=0.$

\noindent{\bf Lemma 2.1}  $\mathcal {R}^{(1)}_{*}<1\Leftrightarrow
y^{(1)}_{1}<y^{(1)}_*$.

\noindent{\bf Proof.}
$$\begin
{array}{lll} \mathcal
{R}^{(1)}_{*}<1&\Leftrightarrow& \frac{(1-\epsilon)\beta s}{ad+(1-\epsilon)a\beta y^{(1)}_*}<1,  \\
&\Leftrightarrow&\mathcal
{R}^{(1)}_{0}<1+\frac{(1-\epsilon)\beta y^{(1)}_*}{d}\\
&\Leftrightarrow& \frac{d(\mathcal
{R}^{(1)}_{0}-1)}{\beta(1-\epsilon)}< y^{(1)}_*\\
&\Leftrightarrow& y^{(1)}_{1}< y^{(1)}_*.
\end {array}$$\qed

 \noindent{\bf Theorem 2.1} ~~If $\mathcal
{R}^{(1)}_{0}<1$, then the uninfected equilibrium $E^{(1)}_{0}$ of
system (1.1) is not only locally asymptotically stable, but also
global asymptotically stable. If $\mathcal {R}^{(1)}_{0}>1$, then
the uninfected equilibrium $E^{(1)}_{0}$ of system (1.1) is
unstable.

\noindent{\bf Proof.}  The characteristic equation of the linearized
system of system (1.1) at $E^{(1)}_0$  is
$$(\lambda+b)(\lambda+d)(\lambda+a-(1-\epsilon)\beta x^{(1)}_{0})=0.$$
Obviously, the characteristic roots  $-d$, $-b$, and $a(\mathcal
{R}^{(1)}_{0}-1)$ are negative for $\mathcal {R}^{(1)}_{0}<1$. Hence
$E^{(1)}_{0}$ is locally asymptotically stable.  If $\mathcal
{R}^{(1)}_{0}>1$, then $a(\mathcal {R}^{(1)}_{0}-1)>0$, thus, the
uninfected equilibrium $E^{(1)}_{0}$ of system (1.1) is unstable.

Consider the Lyapunov function
$$V_0=\frac{1}{2}(x-x^{(1)}_0)^2+x^{(1)}_0 y+\frac{px^{(1)}_0}{M}z.$$
Differentiating $V_0$ along solutions of system (1.1) yields

$$\begin {array}{lll}
\begin {split}
\displaystyle \dot{V_0}|_{(1.1)}&=\displaystyle (x-x^{(1)}_0)[s-dx-(1-\epsilon)\beta xy]+x^{(1)}_0[(1-\epsilon)\beta xy-ay-pyz]\\
&\displaystyle +\frac{px^{(1)}_0}{M}[f(y)z-bz] \vspace{0.2cm}\\
&=\displaystyle (x-x^{(1)}_0)[dx^{(1)}_0-dx-(1-\epsilon)\beta xy]+x^{(1)}_0[(1-\epsilon)\beta xy-ay-pyz]\\
&\displaystyle+\frac{px^{(1)}_0}{M}f(y)z-\frac{px^{(1)}_0}{M}bz  \vspace{0.2cm}\\
&\leq -d(x-x^{(1)}_0)^2-(1-\epsilon)\beta x^2y+2(1-\epsilon)\beta x^{(1)}_0xy-ax^{(1)}_0 y-\frac{px^{(1)}_0}{M}bz\\
&=-[d+(1-\epsilon)\beta y](x-x^{(1)}_0)^2-ax^{(1)}_0
y(1-R^{(1)}_0)-\frac{px^{(1)}_0}{M}bz.
\end{split}
\end {array}$$
If $R^{(1)}_0<1$, then $\dot{V_0}|_{(1.1)}\leq 0$. Furthermore,
$$W_0=\{(x, y, z)| \dot{V_0}=0\}=\{(x, y, z)| x=x^{(1)}_0, y=0, z=0\}.$$
Therefore, the largest invariant set contained in $W_0$ is
$E_0^{(1)}$. By $La Salle's$ invariance principle, \cite{[21],[22]}
we infer that all the solutions of system (1.1) that start in
$R^3>0$ limit to $E_0^{(1)}$. Besides, $E_0^{(1)}$ is Lyapunov
stable, prove that $E_0^{(1)}$ is globally asymptotically stable.
Theorem 2.1 is proved. \qed

\noindent{\bf Theorem 2.2} ~~If $\mathcal {R}^{(1)}_{0}>1>\mathcal
{R}^{(1)}_{*}$, then the immune-free equilibrium $E^{(1)}_{1}$ of
system (1.1) is locally asymptotically stable. $E^{(1)}_{1}$ is
unstable for $\mathcal {R}^{(1)}_{*}>1$.

\noindent{\bf Proof.} The characteristic equation  of the linearized
system of (1.1) at $E^{(1)}_1$ is given by
$$[\lambda-(f(y^{(1)}_{1})-b)][\lambda^2+a^{(1)}_1\lambda+a^{(1)}_2]=0,$$
where
$$\begin
{array}{lll}
a^{(1)}_1&=&d+(1-\epsilon)\beta y^{(1)}_1,\\
a^{(1)}_2&=&(1-\epsilon)^2\beta^2x^{(1)}_1y^{(1)}_1.
\end {array}$$
By (1.2), $f^{'}(y)>0$ for $[0, +\infty)$ and $f(y^{(1)}_*)=b$, we
deduce the eigenvalue $\lambda=f(y^{(1)}_{1})-b<0$ for $\mathcal
{R}^{(1)}_{0}>1>{R}^{(1)}_{*}$, and $\lambda=f(y^{(1)}_{1})-b>0$ for
$\mathcal {R}^{(1)}_{*}>1$. $a^{(1)}_1>0$ and $a^{(1)}_2>0$
inducing, the other eigenvalues are negative. Thus, the immune-free
equilibrium $E^{(1)}_{1}$ of system (1.1) is locally asymptotically
stable for  $\mathcal {R}^{(1)}_{0}>1>\mathcal {R}^{(1)}_{*}$ and
$E^{(1)}_{1}$ is unstable for ${R}^{(1)}_{*}>1$. \qed

\noindent{\bf Theorem 2.3} ~~If $\mathcal {R}^{(1)}_{*}>1$, then the
positive equilibrium $E^{(1)}_*$ of system (1.1) is locally
asymptotically stable.

\noindent{\bf Proof.} The characteristic equation  of the linearized
system of (1.1) at $E^{(1)}_*$ is given by
$$\lambda^3+b^{(1)}_1\lambda^{2}+b^{(1)}_2\lambda+b^{(1)}_{3}=0,$$
where
$$\begin
{array}{lll}
b^{(1)}_1&=&d+(1-\epsilon)\beta y^{(1)}_*,\\
b^{(1)}_2&=&py^{(1)}_*z^{(1)}_*f^{'}(y^{(1)}_*)+(1-\epsilon)^2\beta^2x^{(1)}_*y^{(1)}_*,\\
b^{(1)}_3&=&py^{(1)}_*z^{(1)}_*f^{'}(y^{(1)}_*)[d+(1-\epsilon)\beta
y^{(1)}_*].
\end {array}$$
It is easy to see, $b^{(1)}_i>0 (i=1, 2, 3)$ and
$b^{(1)}_1b^{(1)}_2-b^{(1)}_3=(1-\epsilon)^2\beta^2x^{(1)}_*y^{(1)}_*[d+(1-\epsilon)\beta
y^{(1)}_*]>0 $.  By Routh-Hurartz Criterion, we know the positive
equilibrium $E^{(1)}_*$ of system (1.1) is locally asymptotically
stable for $\mathcal {R}^{(1)}_*>1$. \qed

By Theorem 2.1$\sim$2.3, we can get following result:

\noindent{\bf Remark 2.1 } Viral infection system with monotonic
immune response has no bistability appear.

\section{Viral infection system with nonmonotonic immune response}

\subsection{Equilibria and thresholds}

In this section, we discuss the viral infection system with
nonmonotonic immune response (1.3) and always assume
$\gamma>2\sqrt{\alpha}$. We denote basic reproductive number
${R}^{(2)}_0 =\frac{s\beta (1-\epsilon)}{ad}$, which is equivalent
to ${R}^{(1)}_0$.

 (i) If  $\mathcal
{R}^{(2)}_0<1$, system (1.3) only exists an uninfected equilibrium
$E^{(2)}_{0}=(x^{(2)}_0, 0,0)$ , where $x^{(2)}_0=\frac{s}{d}$ .

 (ii) If  $\mathcal
{R}^{(2)}_0>1$, system (1.3) also has an immune-free equilibrium
$E^{(2)}_{1}=(x^{(2)}_{1}, y^{(2)}_1,0), $ where
$x^{(2)}_{1}=\frac{a}{\beta(1-\epsilon)},y^{(2)}_1=\frac{d({R}^{(2)}_0-1)}{\beta(1-\epsilon)}.$

Solving equation $\frac{cy}{\alpha+\gamma y+y^2}-b=0$, one get two
positive roots, $c_{1}=\gamma b-2b\sqrt{\alpha}$ and $c_{2}=\gamma
b+2b\sqrt{\alpha}$ , then  the existence conditions of positive
equilibria as following:

(iii) If  $\mathcal {R}_{*}^{1-}>1$  and  $c>c_{2},$ system (1.3)
has an immune equilibrium $E_{*}^{2-}=(x_{*}^{2-},y_{*}^{2-},
z_{*}^{2-})$; If $\mathcal {R}_{*}^{1+}>1$ and $c>c_{2},$ system
(1.3) also has an immune equilibrium
$E_{*}^{2+}=(x_{*}^{2+},y_{*}^{2+}, z_{*}^{2+}).$ Here $\mathcal
{R}_{*}^{1\pm}=\frac{(1-\epsilon)\beta s-ad}{(1-\epsilon)\beta a
y_{*}^{2\pm}}, x_{*}^{2\pm}=\frac{s}{ (1-\epsilon)\beta
y_{*}^{2\pm}+d}, y_{*}^{2\pm}=\frac{-B\pm\sqrt{B^{2}-4\alpha
b^2}}{2b}, z_{*}^{2\pm}=\frac{(1-\epsilon)\beta
ay_{*}^{2\pm}(\mathcal {R}_{*}^{1\pm}-1)}{ p [(1-\epsilon)\beta
y_{*}^{2\pm}+d]},  B= \gamma b-c.$

  We denote post-treatment control threshold $P_{I}$ (see e.g. Refs \cite{[19]})
  $$c_{2}=\gamma b+2b\sqrt{\alpha}.$$
  Denote

   $$c^{*}_1=\gamma b+\frac{2bd(\mathcal
{R}^{(2)}_0-1)}{\beta(1-\epsilon)},$$
$$c^{**}_1=\gamma b+\frac{bd(\mathcal
{R}^{(2)}_0-1)}{\beta(1-\epsilon)}+\frac{\alpha \beta
b(1-\epsilon)}{d(\mathcal {R}^{(2)}_0-1)}.$$
We call $c^{**}_1$ the elite control threshold $E_1$, \cite{[19]} which means the virus will be under control when the immune intensity $c$ is larger than $c^{**}_1$.\\
Denote another threshold
 $$\mathcal
{R}^{(1)}_{c}=1+\frac{\beta(1-\epsilon)}{d}\sqrt{\alpha}.$$

For the positive parameters in model (1.3), we have the following
lemmas.

 \noindent{\bf Lemma 3.1 } $\mathcal
{R}^{(2)}_0>\mathcal {R}^{(1)}_{c}>1\Leftrightarrow
c^{*}_1>c^{**}_1.$

\noindent{\bf Proof.}
$$\begin
{array}{lll} c^{*}_1>c^{**}_1&\Leftrightarrow& \frac{bd(\mathcal
{R}^{(2)}_0-1)}{\beta(1-\epsilon)}>\frac{\alpha \beta
b(1-\epsilon)}{d(\mathcal
{R}^{(2)}_0-1)},  \\
&\Leftrightarrow&\mathcal {R}^{(2)}_0>\mathcal {R}^{(1)}_{c}.
\end {array}$$
\qed

\noindent{\bf Lemma 3.2 } (i) $\mathcal {R}^{(2)}_0>\mathcal
{R}^{(1)}_{c}>1\Leftrightarrow c^{*}_1>c_{2}$; (ii) $1<\mathcal
{R}^{(2)}_0<\mathcal {R}^{(1)}_{c}\Leftrightarrow c^{*}_1<c_{2}.$

\noindent{\bf Proof.}
$$\begin
{array}{lll} c^{*}_1>c_{2}&\Leftrightarrow&\frac{2bd(\mathcal
{R}^{(2)}_0-1)}{\beta(1-\epsilon)}>2b\sqrt{\alpha},  \\
&\Leftrightarrow&\mathcal {R}^{(2)}_0>R^{(1)}_{c}.
\end {array}$$
$$\begin
{array}{lll} c^{*}_1<c_{2}&\Leftrightarrow&\frac{2bd({R}^{(2)}_0-1)}{\beta(1-\epsilon)}<2b\sqrt{\alpha},  \\
&\Leftrightarrow&\mathcal {R}^{(2)}_0<R^{(1)}_{c}.
\end {array}$$
\qed

\noindent{\bf Lemma 3.3 } (i) Assume $1<\mathcal
{R}^{(2)}_0<\mathcal {R}^{(1)}_{c}.$ If $\mathcal {R}_{*}^{1-}>1$,
then $c>c^{**}_1$; (ii) Assume $\mathcal {R}^{(2)}_0>\mathcal
{R}^{(1)}_{c}>1.$  If $\mathcal {R}_{*}^{1-}>1$, then $c>c_{2}$.

\noindent{\bf Proof.}
$$\begin
{array}{lll} \mathcal
{R}_{*}^{1-}>1&\Leftrightarrow&\frac{\beta s(1-\epsilon)-ad}{\beta a(1-\epsilon) y_{*}^{2-}}>1,  \\
&\Leftrightarrow&\sqrt{(\gamma b-c)^2-4\alpha b^2}>c-c^{*}_1.
\end {array}$$
If $c<c^{*}_1$ and one of conditions  $c<c_{1}$ or $c>c_{2}$ is
correct, then $\mathcal {R}_{*}^{1-}$ is always larger than one. If
$c>c^{*}_1$, solving $\sqrt{(\gamma b-c)^2-4\alpha b^2}>c-c^{*}_1$,
we have $c>c^{**}_1.$ Thus,

(i) If $1<\mathcal {R}^{(2)}_0<\mathcal {R}^{(1)}_{c}$, then
$c^{*}_1<c_{2}$. From $\mathcal {R}_{*}^{1-}>1$, we have
$c>c^{**}_1.$

(ii) If $\mathcal {R}^{(2)}_{0}>\mathcal {R}^{(1)}_{c}>1$, then
$c^{*}_1>c_{2}$. From $\mathcal {R}_{*}^{1-}>1$, we have  $c>c_{2}.$
\qed

\noindent{\bf Lemma 3.4 } (i) If $1<\mathcal {R}^{(2)}_0<\mathcal
{R}^{(1)}_{c},$ then $\mathcal {R}_{*}^{1+}>1$ has no  solution;
(ii) Assume $\mathcal {R}^{(2)}_0>\mathcal {R}^{(1)}_{c}>1$. If
$\mathcal {R}_{*}^{1+}>1$, then $c_{2}<c<c^{**}_1$.

 \noindent{\bf Proof.}
$$\begin
{array}{lll} \mathcal {R}_{*}^{1+}>1
&\Leftrightarrow& \frac{\beta s(1-\epsilon)-ad}{\beta a(1-\epsilon) y_{*}^{2+}}>1,\\
&\Leftrightarrow&c^{*}_1-c>\sqrt{(\gamma b-c)^2-4\alpha b^2}.
\end {array}$$

(i) If $1<{R}^{(2)}_0<R^{(1)}_{c},$   then $c^{*}_1<c_{2}$. Thus
$\mathcal {R}_{*}^{1+}>1$ has no  solution. (ii) If $\mathcal
{R}^{(2)}_0>\mathcal {R}^{(1)}_{c}>1$, then $c^{*}_1>c_{2}$. Solving
$\mathcal {R}_{*}^{1+}>1$, we have $c_{2}<c<c^{**}_1$. \qed

By Lemma 3.1 $\sim$ Lemma 3.4  and summing up the above analysis we
obtain the existing results of equilibria of system (1.3).

\noindent{\bf Theorem 3.1} (i) System (1.3) always exists an
uninfected equilibrium $E^{(2)}_{0};$

(ii) If  $\mathcal {R}^{(2)}_0>1$, system (1.3) also has an
immune-free equilibrium $E^{(2)}_{1};$

(iii) If $1<\mathcal {R}^{(2)}_0<\mathcal {R}^{(1)}_{c}$ and
$c>c^{**}_1,$ system (1.3) also has  one positive equilibrium
$E_{*}^{2-};$

(iv) If $\mathcal {R}^{(2)}_0>\mathcal {R}^{(1)}_{c}>1$ and
 $c_{2}<c<c^{**}_1$, system (1.3)  has  two positive equilibria
$E_{*}^{2+}$ and  $E_{*}^{2-}$. While  $\mathcal
{R}^{(2)}_0>\mathcal {R}^{(1)}_{c}$ and $c>c^{**}_1$, system (1.3)
only has one positive equilibrium $E_{*}^{2-}$;

The summary results of the existence for positive equilibria can be
seen in Table 1 and Table 2.

\subsection{Stability analysis}
Let $\tilde{E}$ be any arbitrary equilibrium of system (1.3). The
Jacobian matrix associated with the system is

$$J_{2}=\left[
\begin{array}{cccc}
 -d-\beta(1-\epsilon)y  &  -\beta(1-\epsilon)x  & 0 \\
\beta(1-\epsilon)y   & \beta(1-\epsilon)x-a-pz &-py  \\
0   & \frac{c\alpha z-czy^2}{(\alpha+\gamma y+y^2)^2} & \frac{cy}{\alpha +\gamma y+y^2}-b \\
\end{array}
\right].$$ The characteristic equation of the linearized system of
(1.3) at $\tilde{E}$ is given by $\left|\lambda I-J_{2} \right|=0.$

\noindent{\bf Theorem 3.2 } \hspace{0.1cm}   If $\mathcal
{R}^{(2)}_0<1$, then the uninfected equilibrium $E^{(2)}_{0}$ of
system (1.3) is not only locally asymptotically stable, but also
global asymptotically stable.

\noindent{\bf Proof.}  The characteristic roots  of the linearized
system of (1.3) at $E^{(2)}_{0}$ is given by $\lambda_1=-d$,
$\lambda_2=-b$ and  $\lambda_3=\frac{\mathcal {R}^{(2)}_0-1}{a}.$ So
we can get $\mathcal {R}^{(2)}_0<1$ , the uninfected equilibrium
$E^{(2)}_{0}$ is locally asymptotically stable.

Consider the Lyapunov function
$$V_1=\frac{1}{2}(x-x^{(2)}_0)^2+x^{(2)}_0 y+\frac{\alpha px^{(2)}_0}{c}z.$$
Differentiating $V_1$ along solutions of system (1.3) yields

$$\begin {array}{lll}
\begin {split}
\displaystyle \dot{V_1}|_{(1.3)}&=\displaystyle (x-x^{(2)}_0)[s-dx-(1-\epsilon)\beta xy]+x^{(2)}_0[(1-\epsilon)\beta xy-ay-pyz]\\
&\displaystyle +\frac{\alpha px^{(2)}_0}{c}[\frac{cyz}{\alpha+\gamma y+y^2}-bz] \vspace{0.2cm}\\
&=\displaystyle (x-x^{(2)}_0)[dx^{(2)}_0-dx-(1-\epsilon)\beta xy]+x^{(2)}_0[(1-\epsilon)\beta xy-ay-pyz]\\
&\displaystyle+\frac{\alpha px^{(2)}_0yz}{\alpha+\gamma y+y^2}-\frac{\alpha bpx^{(2)}_0}{c}z  \vspace{0.2cm}\\
&\leq -d(x-x^{(2)}_0)^2-(1-\epsilon)\beta x^2y+2(1-\epsilon)\beta x^{(2)}_0xy-ax^{(2)}_0 y-\frac{\alpha bpx^{(2)}_0}{c}z\\
&=-[d+(1-\epsilon)\beta y](x-x^{(2)}_0)^2-ax^{(2)}_0
y(1-R^{(2)}_0)-\frac{\alpha bpx^{(2)}_0}{c}z.
\end{split}
\end {array}$$
If $R^{(2)}_0<1$, then $\dot{V_1}|_{(1.3)}\leq 0$.

Furthermore,
$$W_1=\{(x, y, z)| \dot{V_1}=0\}=\{(x, y, z)| x=x^{(2)}_0, y=0, z=0\}.$$
Therefore, the largest invariant set contained in $W_1$ is
$E_0^{(2)}$. By $La Salle's$ invariance principle, \cite{[21],[22]}
we infer that all the solutions of system (1.3) that start in
$R^3>0$ limit to $E_0^{(2)}$. Besides, $E_0^{(2)}$ is Lyapunov
stable, prove that $E_0^{(2)}$ is globally asymptotically stable.
Theorem 3.2 is proved. \qed

\noindent{\bf Theorem 3.3}  Suppose $\mathcal {R}^{(2)}_0>1$. When
$0<c<c^{**}_1,$ $E^{(2)}_{1}$ is locally asymptotically stable. When
$c>c^{**}_1,$ $E^{(2)}_{1}$ is unstable.

\noindent{\bf Proof.}  The characteristic equation  of the
linearized system of (1.3) at $E^{(2)}_1$ is given by
$[\lambda-(\frac{cy^{(2)}_1}{\alpha +\gamma y^{(2)}_1+{(y^{(2)}_1})^2}-b)][\lambda^2+a^{(2)}_1\lambda+a^{(2)}_2]=0$,\\
where
$$\begin
{array}{lll}
a^{(2)}_1&=&(1-\epsilon)\beta y^{(2)}_1+d>0,\\
a^{(2)}_2&=&(1-\epsilon)^2\beta^2x^{(2)}_1y^{(2)}_1>0.
\end {array}$$
Another  eigenvalue
$$\begin
{array}{lll} \lambda=\frac{cy^{(2)}_1}{\alpha +\gamma
y^{(2)}_1+{(y^{(2)}_1})^2}-b<0 &\Leftrightarrow& c<c^{**}_1.
\end {array}$$

In summary, if $0<c<c^{**}_1,$ then $\lambda<0.$ Therefore, by
Routh-Hurartz criterion, we know under the assumption of  $\mathcal
{R}^{(2)}_0>1$. If $0<c<c^{**}_1,$ the equilibrium $E^{(2)}_{1}$ of
system (1.3) is locally asymptotically stable. If $c>c^{**}_1,$
$E^{(2)}_{1}$ is unstable. \qed

 \noindent{\bf Theorem 3.4}
(i) If \hspace{0.1cm} ($\mathbf{A.1}$) \hspace{0.2cm} $1<\mathcal
{R}^{(2)}_0<\mathcal {R}^{(1)}_{c}$ and $c>c^{**}_1$, or

\hspace{2.3cm} ($\mathbf{A.2}$) \hspace{0.2cm}    $\mathcal
{R}^{(2)}_0>\mathcal {R}^{(1)}_{c}$ and $c>c_{2}$, \\
system (1.3) has  an immune equilibrium $E_{*}^{2-},$ which is a
stable node.

(ii) If  $\mathcal {R}^{(2)}_0>\mathcal {R}^{(1)}_{c}$ and
 $c_{2}<c<c^{**}_1$, system (1.3)
also has an immune equilibrium $E_{*}^{2+},$ which is an unstable
saddle.

\noindent{\bf Proof.} Denote $E_{*}^{(2)}=(x_{*}^{(2)},y_{*}^{(2)},
z_{*}^{(2)})$ as an arbitrary positive equilibrium of system (1.3).
The characteristic equation of the linearized system of (1.3) at the
arbitrary positive equilibrium $E_{*}^{(2)}$ is given by
$$\lambda^3+b^{(2)}_1\lambda^2+b^{(2)}_2\lambda+b^{(2)}_3=0,$$ where
 $$\begin
{array}{lll}
b^{(2)}_1&=&(1-\epsilon)\beta y_{*}^{(2)}+d>0,\\
b^{(2)}_2&=&(1-\epsilon)^2\beta^2x_{*}^{(2)}y_{*}^{(2)}+cpy_{*}^{(2)}z_{*}^{(2)}\frac{\alpha-(y_{*}^{(2)})^2}{(\alpha+\gamma y_{*}^{(2)}+(y_{*}^{(2)})^2)^2},\\
b^{(2)}_3&=&cpy_{*}^{(2)}z_{*}^{(2)}[(1-\epsilon)\beta
y_{*}^{(2)}+d]\frac{\alpha-(y_{*}^{(2)})^2}{(\alpha+\gamma
y_{*}^{(2)}+(y_{*}^{(2)})^2)^2},
\end {array}$$
and
$$b^{(2)}_1b^{(2)}_2-b^{(2)}_3=[(1-\epsilon)\beta
y_{*}^{(2)}+d](1-\epsilon)^2\beta^2x_{*}^{(2)}y_{*}^{(2)}>0.$$

For equilibrium $E_{*}^{2-},$
$$\begin
{array}{lll} \alpha-(y_{*}^{2-})^2>0
&\Leftrightarrow&\frac{-B-\sqrt{B^2-4\alpha b^2}}{2b}<\sqrt{\alpha} ,\\
&\Leftrightarrow&c>c_2.
\end {array}$$
If $c>c_2$, we can get $b^{(2)}_2>0$ and $b^{(2)}_3>0$, by
Routh-Hurartz Criterion, we know in this case the positive
equilibrium $E_{*}^{2-}$ is a stable node.

For equilibrium $E_{*}^{2+},$
$$\begin
{array}{lll} \alpha-(y_{*}^{2+})^2<0
&\Leftrightarrow&\frac{-B+\sqrt{B^2-4\alpha b^2}}{2b}>\sqrt{\alpha} ,\\
&\Leftrightarrow& \sqrt{B^2-4\alpha b^2}>B+2b\sqrt{\alpha}.
\end {array}$$
When $c_2<c<c^{**}_1$, then $b^{(2)}_3<0$, so the immune equilibrium
$E_{*}^{2+}$ is an unstable saddle. \qed

\subsection{Saddle-node bifurcation}
If $\mathcal {R}^{(2)}_{0}>\mathcal {R}^{(1)}_{c}>1$ and
$c^2-2\gamma bc+\gamma^2b^2-4\alpha b^2=0$, the immune equilibrium
$E_{*}^{2+}$ and  $E_{*}^{2-}$ coincide with each other. Then system
has the unique interior equilibrium $E_{*}^{(2)}=(x_{*}^{(2)},
y_{*}^{(2)}, z_{*}^{(2)})=(\frac{s}{(1-\epsilon)\beta \alpha+d},
\sqrt{\alpha}, \frac{1}{p}(\frac{(1-\epsilon)\beta
s}{(1-\epsilon)\beta \alpha+d}-a)$. If $c<c^{[sn]}$, there is no
positive equilibrium and there is two positive equilibria. Thus,
system (1.3) will be a saddle-node bifurcation when $c$ crosses the
bifurcation value $c^{[sn]}$, where $c^{[sn]}=\gamma
b+2b\sqrt{\alpha}$.

\noindent{\bf Theorem 3.5 }If $\mathcal {R}^{(2)}_{0}>\mathcal
{R}^{(1)}_{c}>1$ and $c=c^{[sn]}$, system (1.3) undergoes a
saddle-node bifurcation.

\noindent{\bf Proof.} We use Sotomayor's theorem
\cite{[27],[28],[29]} to prove system (1.3) undergoes a saddle-node
bifurcation at $c=c^{[sn]}$. It can be easy to prove
$Det[J_{E_{*}^{(2)}}]=0$, so one of the eigenvalue of the Jacobian
at the saddle-node equilibrium is zero, where $J=J_2$.

Let $\varphi=(\varphi_1, \varphi_2, \varphi_3)^\mathrm{T}$ and
$\psi=(\psi_1, \psi_2, \psi_3)^\mathrm{T}$ represent the
eigenvectors of $J_{E_{*}^{(2)}}$ and $J_{E_{*}^{(2)}}^\mathrm{T}$
corresponding to the zero eigenvalue, respectively, then they are
given by $\varphi=(1,
\frac{-d-\beta(1-\epsilon)y_{*}^{(2)}}{\beta(1-\epsilon)x_{*}^{(2)}},
\frac{\beta(1-\epsilon)}{p})^\mathrm{T}$ and $\psi=(0, 0,
1)^\mathrm{T}$. Let $G=(g_1, g_2, g_3)$, we can get
$$G_c(E_{*}^{(2)};c^{[sn]})=\left [
\begin{array}{cccc}
0\\
0\\
\frac{yz}{\alpha+\gamma y+y^2}
\end{array}
\right ]_{(E_{*}^{(2)};c^{[sn]})}= \left[
\begin{array}{cccc}
0 \\
0\\
\frac{\sqrt{\alpha}z_{*}^{(2)}}{2\alpha+\gamma \sqrt{\alpha}}
\end{array}
\right ],
$$

$$\begin {array}{lll}
\begin {split}
\displaystyle & D^2G(E_{*}^{(2)};c^{[sn]})(\varphi,\varphi)\\ &=
\left [
\begin{array}{cccc}
\frac{2(d+\beta(1-\epsilon)y)}{x}\\
0\\
\frac{(-6c\alpha yz+2czy^3-2c\alpha\gamma
z)(d+\beta(1-\epsilon)y)^2}{\beta^2(1-\epsilon)^2x^2(\alpha+\gamma
y+y^2)^3}-\frac{2c\beta(1-\epsilon)(d+\beta(1-\epsilon)y)(\alpha-y^2)}{px\beta(1-\epsilon)(\alpha+\gamma
y+y^2)^2}
\end{array}
\right ]_{(E_*;c^{[sn]})}\\
&=\left[
\begin{array}{cccc}
\frac{2(d+\beta(1-\epsilon)\sqrt{\alpha})}{x_{*}^{(2)}} \\
0\\
\frac{2\alpha z_{*}^{(2)}(\gamma b+2b\sqrt{\alpha})(d+\beta
\sqrt{\alpha}(1-\epsilon))^2}{\beta^2{x_{*}^{(2)}}^2(1-\epsilon)^2(2\alpha+\gamma
\sqrt{\alpha})^3}
\end{array}
\right ].
\end{split}
\end {array}$$

Therefore,
$$\begin {array}{lll}
\begin {split}
\Psi_1&=\psi^\mathrm{T}G_c(E_{*}^{(2)},c^{[sn]})=\frac{\sqrt{\alpha}z_{*}^{(2)}}{2\alpha+\gamma \sqrt{\alpha}}\neq0,   \vspace{0.2cm}\\
\Psi_2&=\psi^\mathrm{T}[D^2G(E_{*}^{(2)};c^{[sn]})(\varphi,
\varphi)]=\frac{2\alpha z_{*}^{(2)}(\gamma
b+2b\sqrt{\alpha})(d+\beta
\sqrt{\alpha}(1-\epsilon))^2}{\beta^2{x_{*}^{(2)}}^2(1-\epsilon)^2(2\alpha+\gamma
\sqrt{\alpha})^3}\neq0.
\end{split}
\end {array}$$

Therefore, system (1.3) undergoes a saddle-node bifurcation at
$E_*^{(2)}$ when $c=c^{[sn]}$. If $c<c^{[sn]}$, there is no positive
equilibrium. If $c>c^{[sn]}$, there is two positive equilibria.

\subsection{Transcritical Bifurcation }

If $c=\gamma b+\frac{bd(\mathcal
{R}^{(2)}_0-1)}{1-\epsilon}+\frac{\alpha \beta
b(1-\epsilon)}{d(\mathcal {R}^{(2)}_0-1)}$, the boundary equilibrium
$E_1^{(2)}$ looses its stability and one of the eigenvalue of the
Jacobian at $E_1^{(2)}$ is zero. Hence, bifurcation may occur at the
boundary equilibrium $E_1^{(2)}$. Next we study the existence of a
transcritical bifurcation and select parameter $c$ as bifurcation
parameter.

\noindent{\bf Theorem 5.6 } If $R_0>1$ and $c=c^{[tc]}$, system
(1.3) will undergoes a transcritical bifurcation at $E_1^{(2)}$, $c$
as the bifurcation parameter and $c^{[tc]}$ as the bifurcation
threshold is given by $c=c^{[tc]}=\gamma b+\frac{bd(\mathcal
{R}^{(2)}_0-1)}{\beta(1-\epsilon)}+\frac{\alpha \beta
b(1-\epsilon)}{d(\mathcal {R}^{(2)}_0-1)}$.

\noindent{\bf Proof.} We also use Sotomayor's theorem
\cite{[27],[28],[29]} to prove system (1.3) undergoes a
transcritical bifurcation. It is clear that one of the eigenvalue of
the Jacobian at $E_1^{(2)}$ is zero, if and only if $c=c^{[tc]}$.

Let $\eta=(\eta_1, \eta_2, \eta_3)^\mathrm{T}$ and
$\theta=(\theta_1, \theta_2, \theta_3)^\mathrm{T}$ denote the
eigenvectors of $J_{E_1^{(2)}}$ and $J_{E_1^{(2)}}^\mathrm{T}$
corresponding to the zero eigenvalue, respectively, we can get $\eta
=(1,
\frac{-d-\beta(1-\epsilon)y_{1}^{(2)}}{\beta(1-\epsilon)x_{1}^{(2)}},
\frac{\beta(1-\epsilon)}{p})^\mathrm{T}$ and $\theta=(0, 0,
1)^\mathrm{T}$, Besides,
$$G_c(E_1^{(2)};c^{[tc]})=\left [
\begin{array}{cccc}
0\\
0\\
\frac{yz}{\alpha+\gamma y+y^2}
\end{array}
\right ]_{(E_1^{(2)};c^{[tc]})}= \left[ \begin{array}{cccc}
0 \\
0\\
0
\end{array}
\right ].
$$
$$\begin {array}{lll}
\begin {split}
\displaystyle & DG_c(E_1^{(2)};c^{[tc]})\eta\\ &= \left [
\begin{array}{cccc}
0 \\
0 \\
-\frac{(d+\beta(1-\epsilon)y)(\alpha-y^2)z}{\beta
x(1-\epsilon)(\alpha+\gamma y+y^2)^2}+ \frac{\beta
y(1-\epsilon)}{p(\alpha+\gamma y+y^2)}
\end{array}
\right ]_{(E_1^{(2)};c^{[tc]})}\\
&=\left[
\begin{array}{cccc}
0 \\
0\\
\frac{\beta y_1^{(2)}(1-\epsilon)}{p(\alpha+\gamma
y_1^{(2)}+{y_1^{(2)}}^2)}
\end{array}
\right ].
\end{split}
\end {array}$$
$$\begin {array}{lll}
\begin {split}
\displaystyle & D^2G(E_{1}^{(2)};c^{[sn]})(\eta,\eta)\\ &= \left [
\begin{array}{cccc}
\frac{2(d+\beta(1-\epsilon)y)}{x}\\
0\\
\frac{(-6c\alpha yz+2czy^3-2c\alpha\gamma
z)(d+\beta(1-\epsilon)y)^2}{\beta^2(1-\epsilon)^2x^2(\alpha+\gamma
y+y^2)^3}-\frac{2c\beta(1-\epsilon)(d+\beta(1-\epsilon)y)(\alpha-y^2)}{px\beta(1-\epsilon)(\alpha+\gamma
y+y^2)^2}
\end{array}
\right ]_{(E_{1}^{(2)};c^{[sn]})}\\
&=\left[
\begin{array}{cccc}
\frac{2(d+\beta(1-\epsilon)\sqrt{\alpha})}{x_{1}^{(2)}} \\
0\\
-\frac{2c\beta(1-\epsilon)(d+\beta(1-\epsilon)y_{1}^{(2)})(\alpha-{y_{1}^{(2)}}^2)}{px_{1}^{(2)}\beta(1-\epsilon)(\alpha+\gamma
y_{1}^{(2)}+{y_{1}^{(2)}}^2)^2}
\end{array}
\right ].
\end{split}
\end {array}$$
Therefore,
$$\begin {array}{lll}
\begin {split}
\Gamma_1&=\theta^\mathrm{T}G_c(E_1^{(2)};c^{[tc]})=0,   \vspace{0.2cm}\\
\Gamma_2&=\theta^\mathrm{T}[DG_c(E_1^{(2)};c^{[tc]})\eta]=\frac{\beta y_1^{(2)}(1-\epsilon)}{p(\alpha+\gamma y_1^{(2)}+{y_1^{(2)}}^2)}\neq0\\
\Gamma_3&=\theta^\mathrm{T}[D^2G(E_1^{(2)};c^{[tc]})(\eta,\eta)]=-\frac{2c\beta(1-\epsilon)
(d+\beta(1-\epsilon)y_{1}^{(2)})(\alpha-{y_{1}^{(2)}}^2)}{px_{1}^{(2)}\beta(1-\epsilon)(\alpha+\gamma
y_{1}^{(2)}+{y_{1}^{(2)}}^2)^2}\neq0.
\end{split}
\end {array}$$

Therefore, system (1.3) will undergoes a transcritical bifurcation
between $E_1^{(2)}$ when $c=c^{[tc]}$

\qed

 \noindent{\bf Remark 3.1 }   If $\mathcal
{R}^{(2)}_{0}>\mathcal {R}^{(1)}_{c}>1$ and
 $c_{2}<c<c^{**}_1$, system (1.3) has bistability appear. In other cases, system (1.3) has no bistability appear. Threshold $c_{2}$ is a post-treatment control threshold, $c^{**}_1$ is a elite control threshold. $(c_{2}, c^{**}_1)$ is a bistable interval.

To sum up, the stabilities of the equilibria and the behaviors of
system (1.3) can be shown in Table 3 and Table 4.

\subsection{Numerical simulations and discussion}
To verify our analysis results,  we carry out some numerical
simulations choosing some parameter values shown as in
\cite{[19],[23],[24]}:
$$\begin
{array}{lll} &&s=10~~cells/\mu l/day ,\hspace{0.02cm}  d=0.01~~\mbox{day}^{-1}, \hspace{0.02cm}  \epsilon=0.9,\\
&& \beta=0.015~~\mu l/day ,\hspace{0.02cm}  a=1.1~~\mbox{day}^{-1},\hspace{0.02cm}  p=0.5 ~~\mbox{day}^{-1},\\
&& \alpha=1 ~~cells/\mu l,\hspace{0.02cm}  \gamma=1 ~~cells/\mu
l,\hspace{0.02cm}  b=0.1~~\mbox{day}^{-1}.
\end {array}
   \eqno(3.1)$$

The parameters chose as same as in (3.1), the thresholds $\mathcal
{R}^{(2)}_{0}\approx 1.3636$, $\mathcal {R}^{(1)}_{c}= 1.1500,$
post-treatment control threshold $c_{2}=0.3000$ and elite control
threshold $c^{**}_1\approx0.3837$.  In this case, $\mathcal
{R}^{(2)}_{0}>\mathcal {R}^{(1)}_{c}$ and $c_{2}<c^{**}_1$, then we
get a bistable interval $(0.3000, 0.3837)$ (see Figure 1). When
$0<c<c_{2}$, the immune-free equilibrium $E^{(2)}_1$ is stable (see
Fig. 2); When $c_{2}<c<c^{**}_1$, the immune-free equilibrium
$E^{(2)}_1$ and the positive equilibrium $E_{*}^{2-}$ are stable
(see Fig. 3); When $c>c^{**}_1$, only the positive equilibrium
$E_{*}^{2-}$ is stable (see Figure 4).


\begin{figure}[!h]
\begin{center}
{\rotatebox{0}{\includegraphics[width=0.7 \textwidth,
height=50mm]{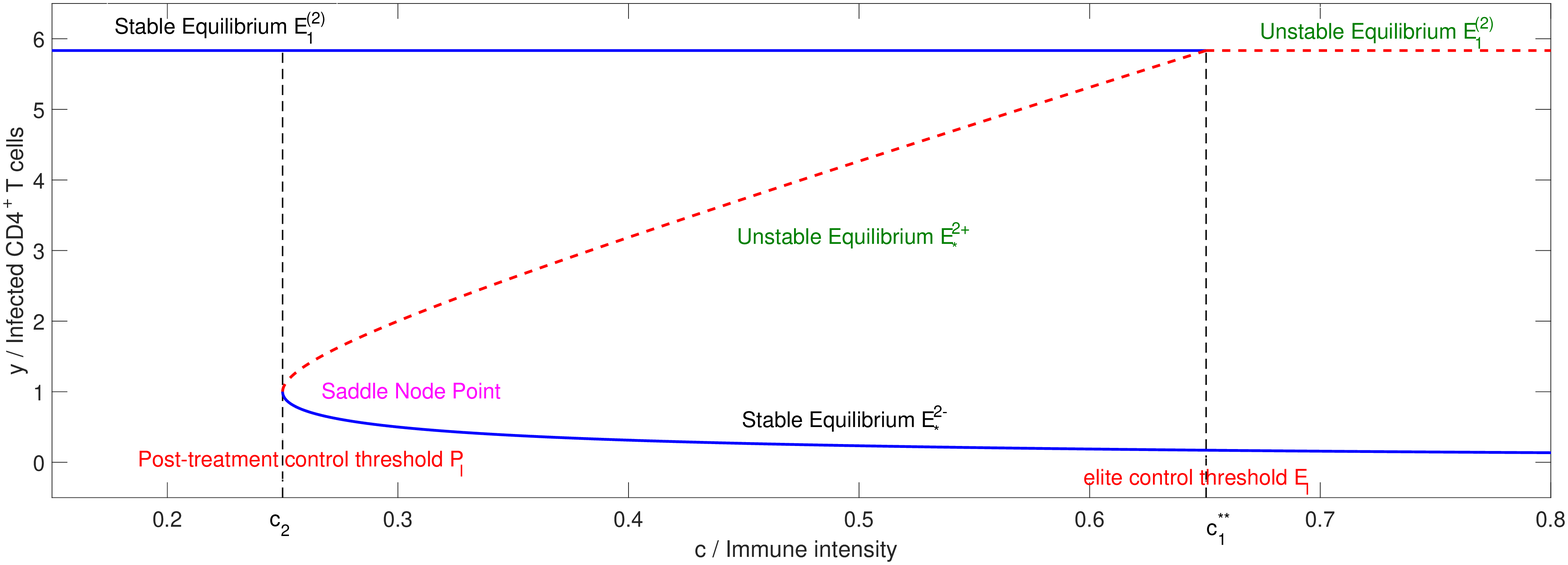}}}
 \caption{
\footnotesize Bistability and saddle-node bifurcation diagram of
system (1.3). The solid line is the stable infected CD4+ T cells and
the dashed line depends the unstable infected CD4+ T cells. The
post-treatment control threshold is $c_{2}=0.2500$, the elite
control threshold is $c^{**}_1\approx0.6505$ and the bistable
interval is $(0.2500, 0.6505).$ $c=0.37~~\mbox{day}^{-1}$ and other
parameter values are shown in (3.1). }\label{F51}
\end{center}
 \end{figure}

\begin{figure}[!h]
\begin{center}
{\rotatebox{0}{\includegraphics[width=0.4 \textwidth,
height=40mm]{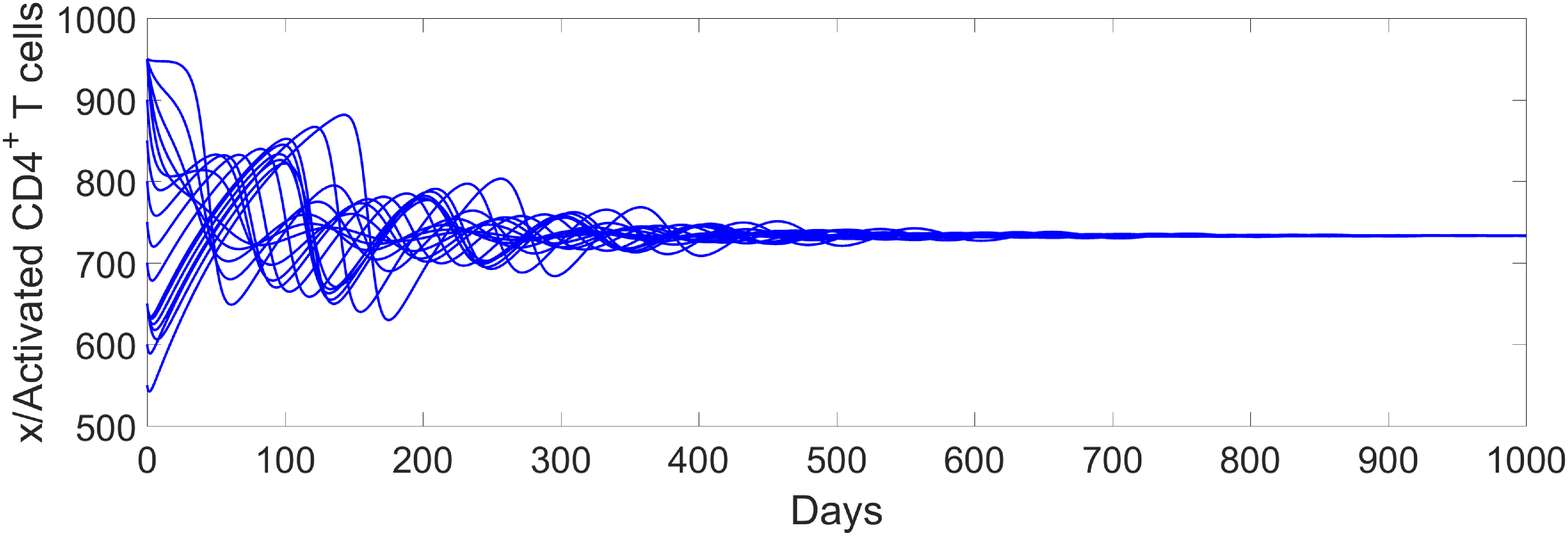}}} {\rotatebox{0}{\includegraphics[width=0.4
\textwidth, height=40mm]{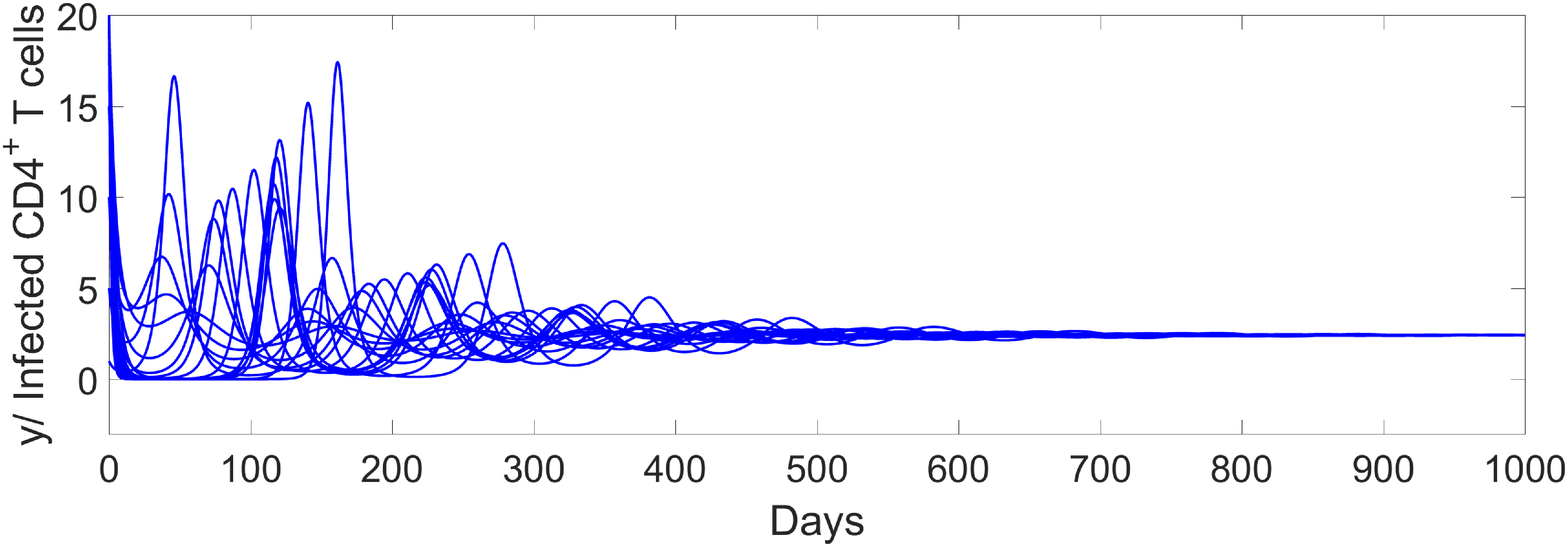}}}\\
{\rotatebox{0}{\includegraphics[width=0.4 \textwidth,
height=40mm]{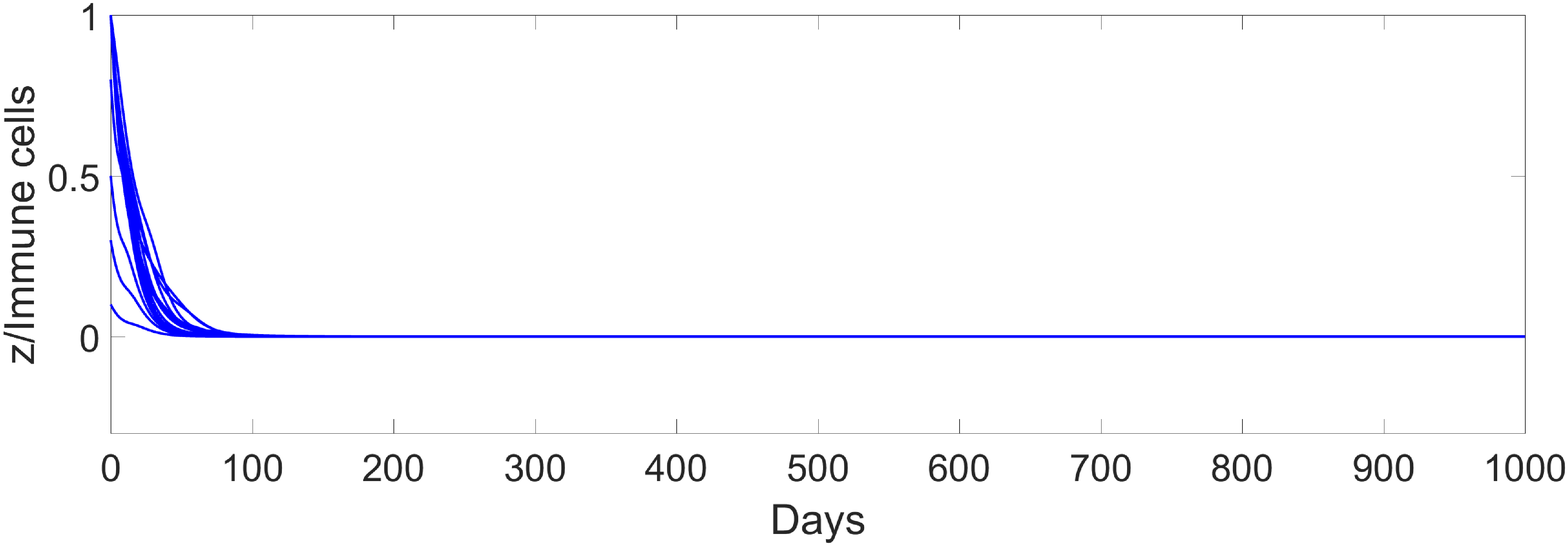}}} {\rotatebox{0}{\includegraphics[width=0.4
\textwidth, height=45mm]{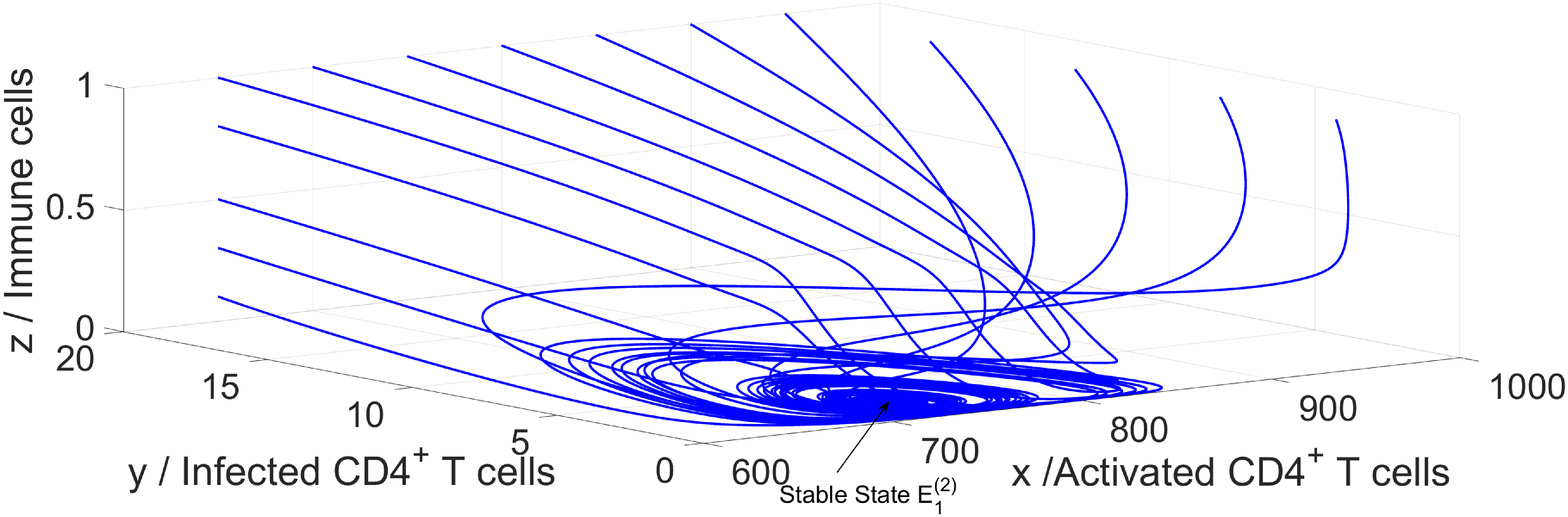}}}
 \caption{
\footnotesize System (1.3) has a stable equilibria $E^{(2)}_{1}$.
Parameter $c=0.2~~\mbox{day}^{-1}$ less than post-treatment control
threshold $P_I$ and other parameter values are shown in (3.1). We
choose different initial values. }\label{F51}
\end{center}
 \end{figure}
\begin{figure}[!h]
\begin{center}
{\rotatebox{0}{\includegraphics[width=0.4 \textwidth,
height=40mm]{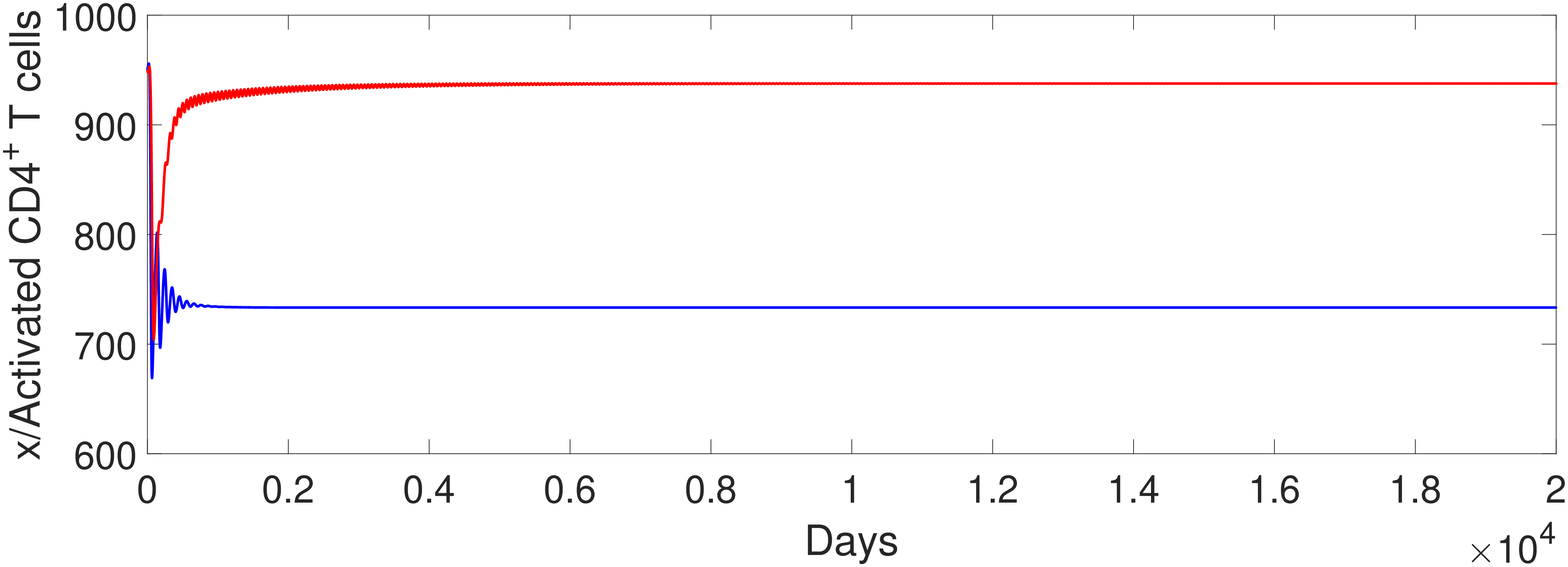}}} {\rotatebox{0}{\includegraphics[width=0.4
\textwidth, height=40mm]{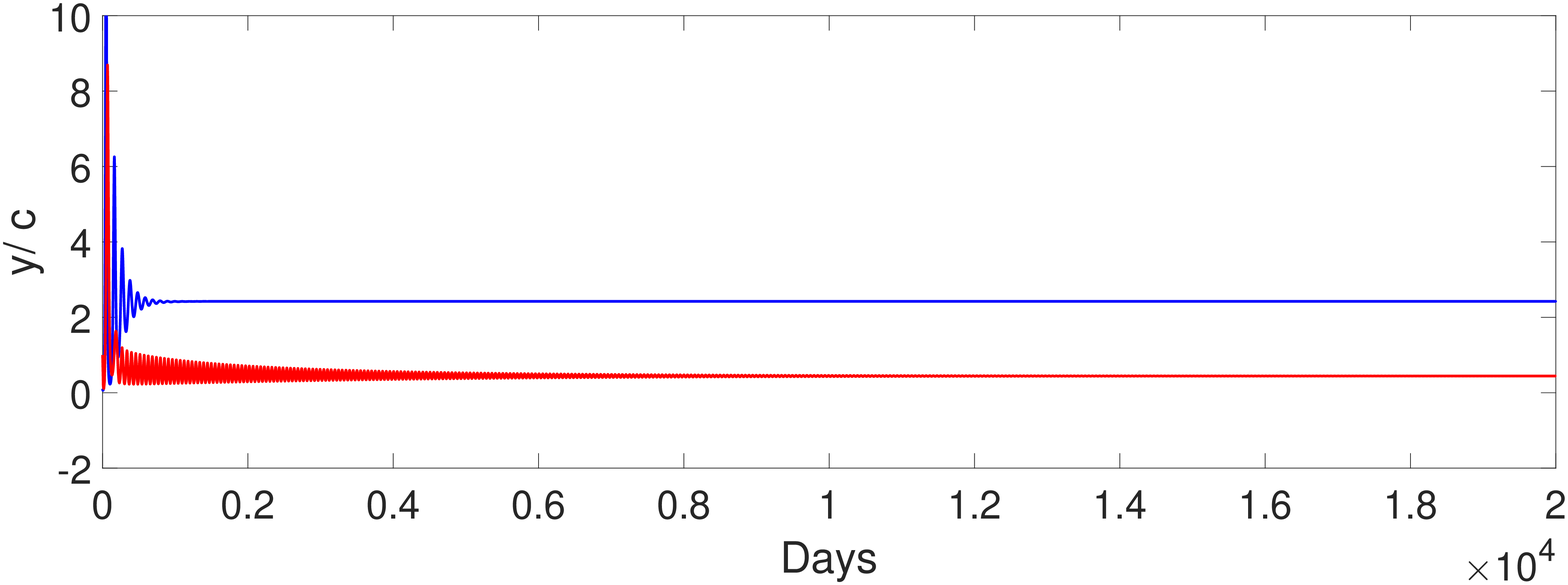}}}\\
{\rotatebox{0}{\includegraphics[width=0.4 \textwidth,
height=40mm]{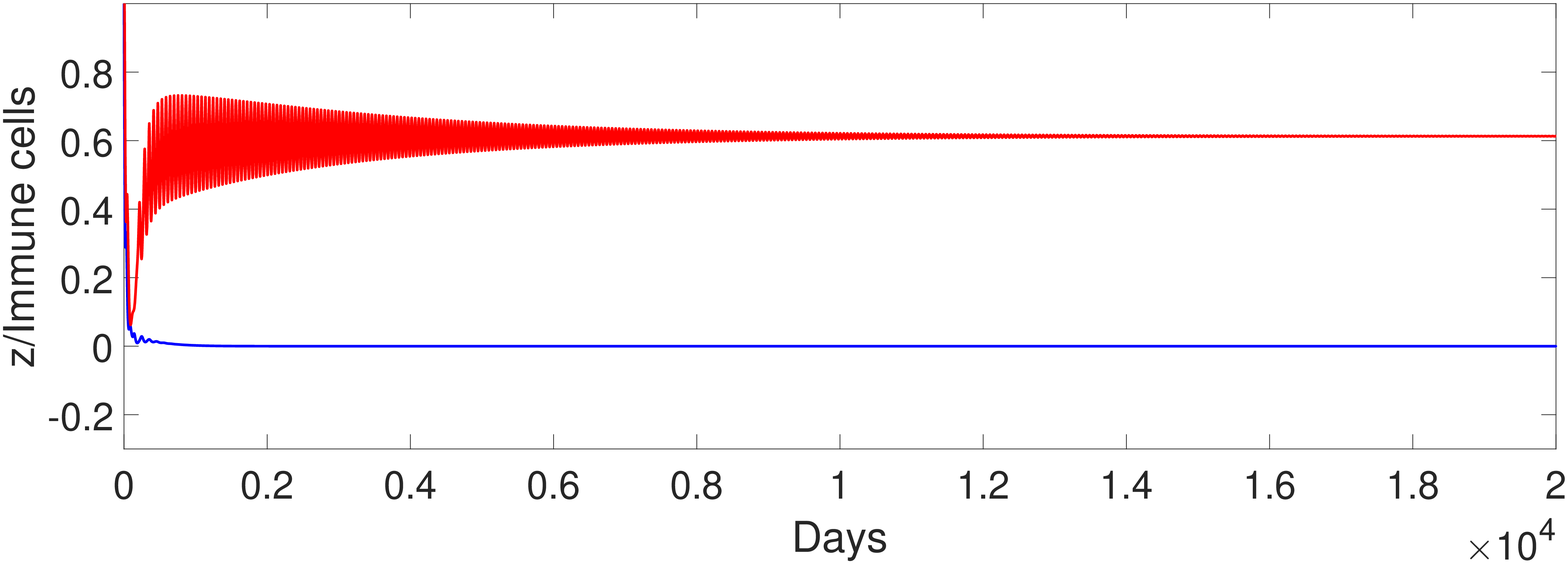}}} {\rotatebox{0}{\includegraphics[width=0.4
\textwidth, height=45mm]{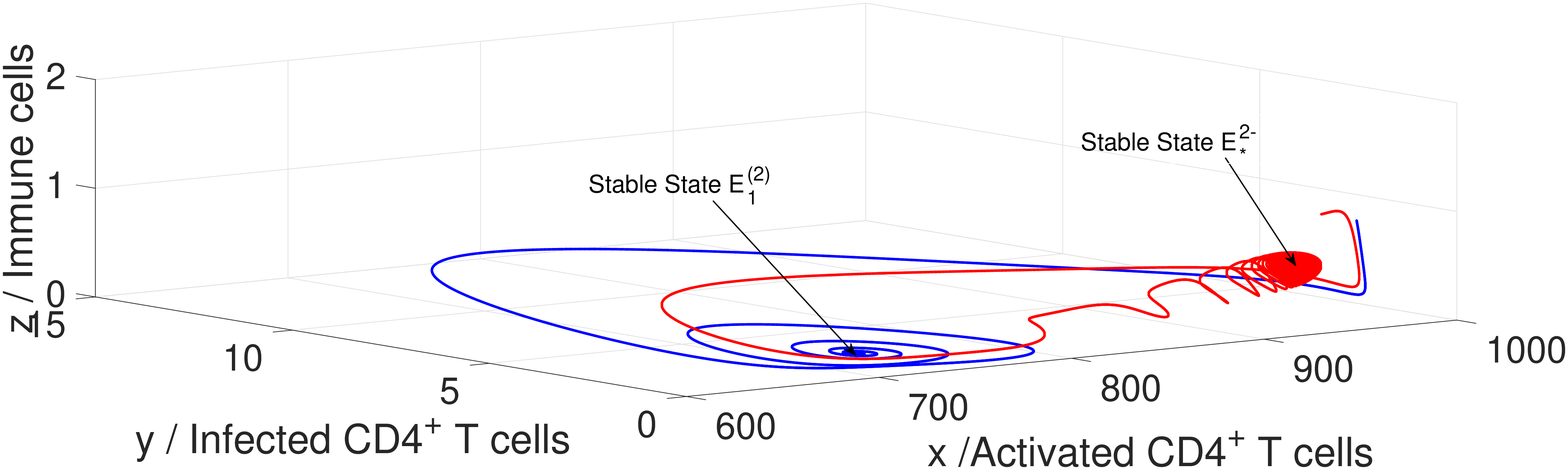}}}
 \caption{
\footnotesize System (1.3) has two different stable equilibria
$E^{(2)}_{1}$ and $E_{-}^{2*}$. Parameter $c=0.37~~\mbox{day}^{-1}$
and other parameter values are shown in (3.1). We choose different
initial values. }\label{F51}
\end{center}
 \end{figure}
\begin{figure}[!h]
\begin{center}
{\rotatebox{0}{\includegraphics[width=0.4 \textwidth,
height=40mm]{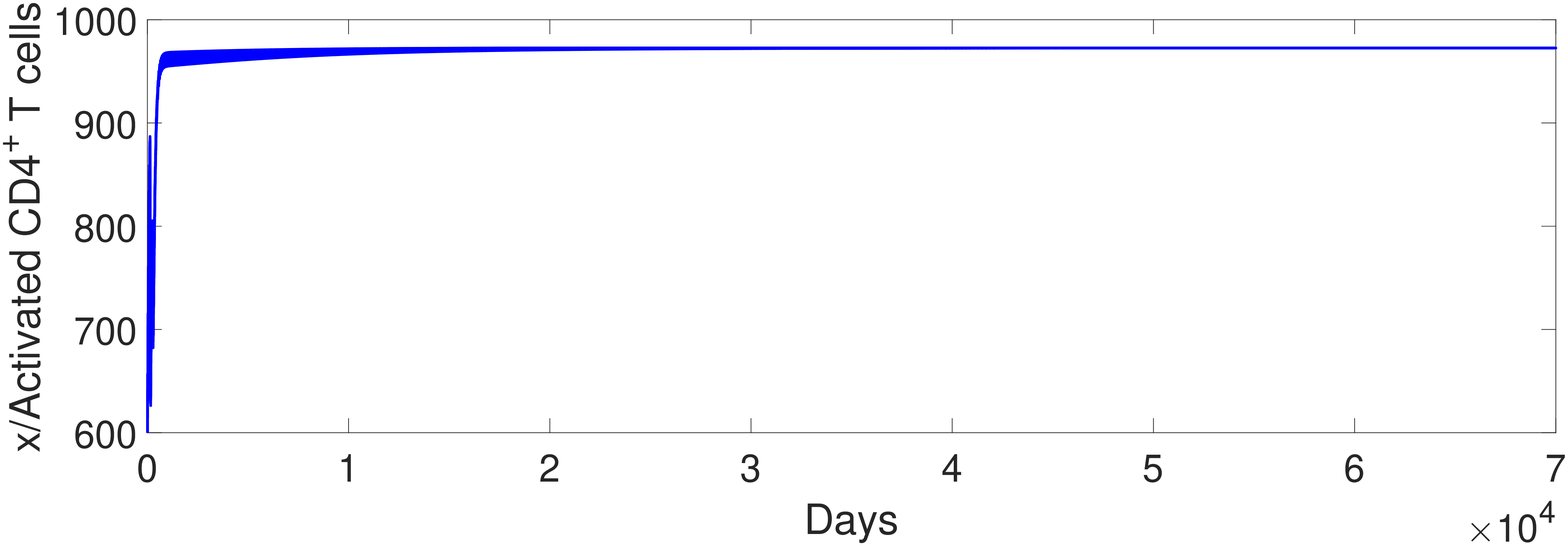}}} {\rotatebox{0}{\includegraphics[width=0.4
\textwidth, height=40mm]{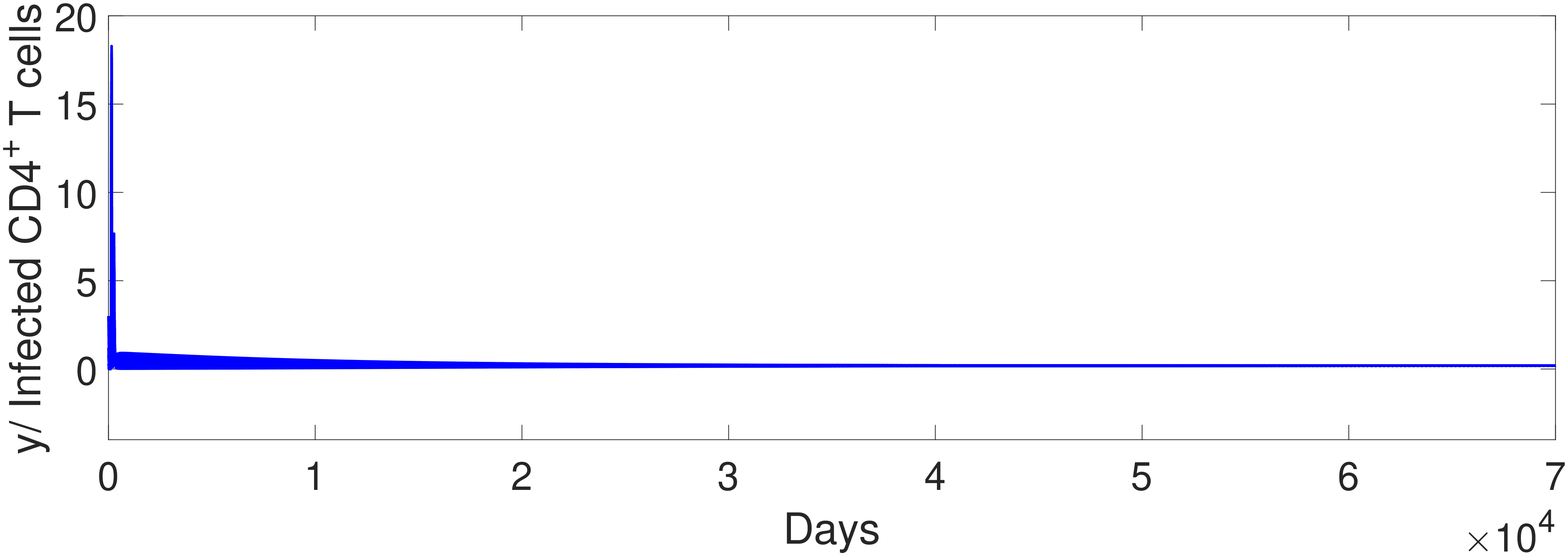}}}\\
{\rotatebox{0}{\includegraphics[width=0.4 \textwidth,
height=40mm]{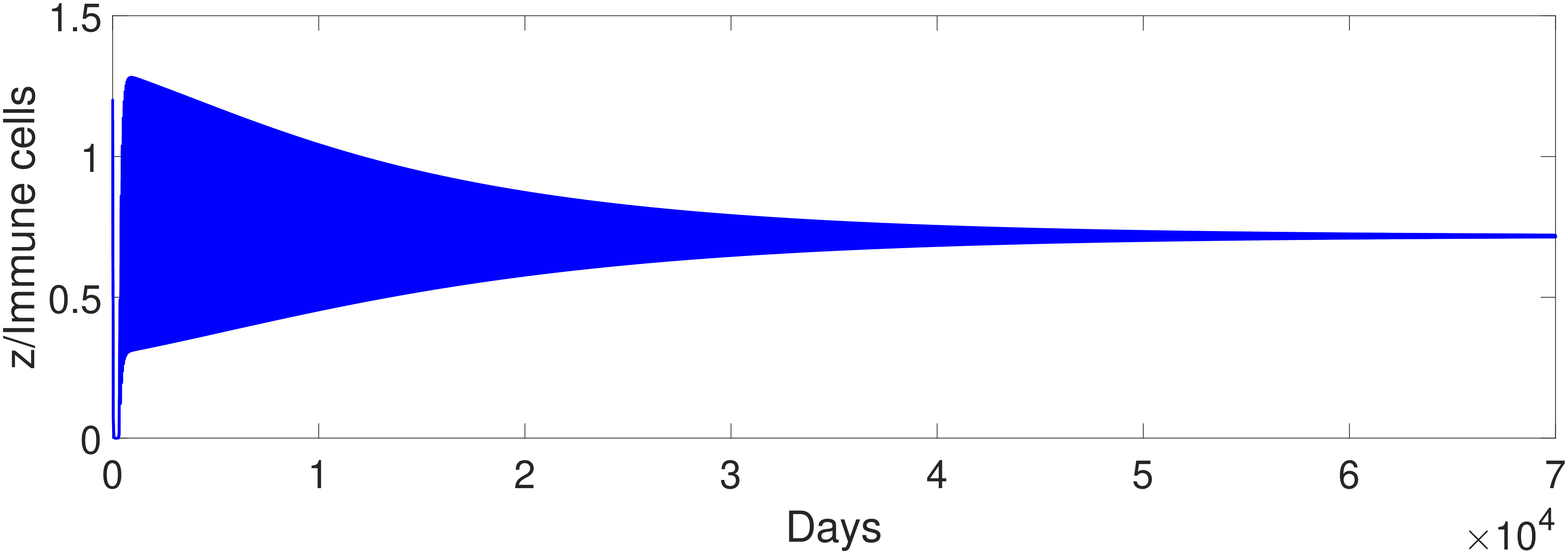}}} {\rotatebox{0}{\includegraphics[width=0.4
\textwidth, height=45mm]{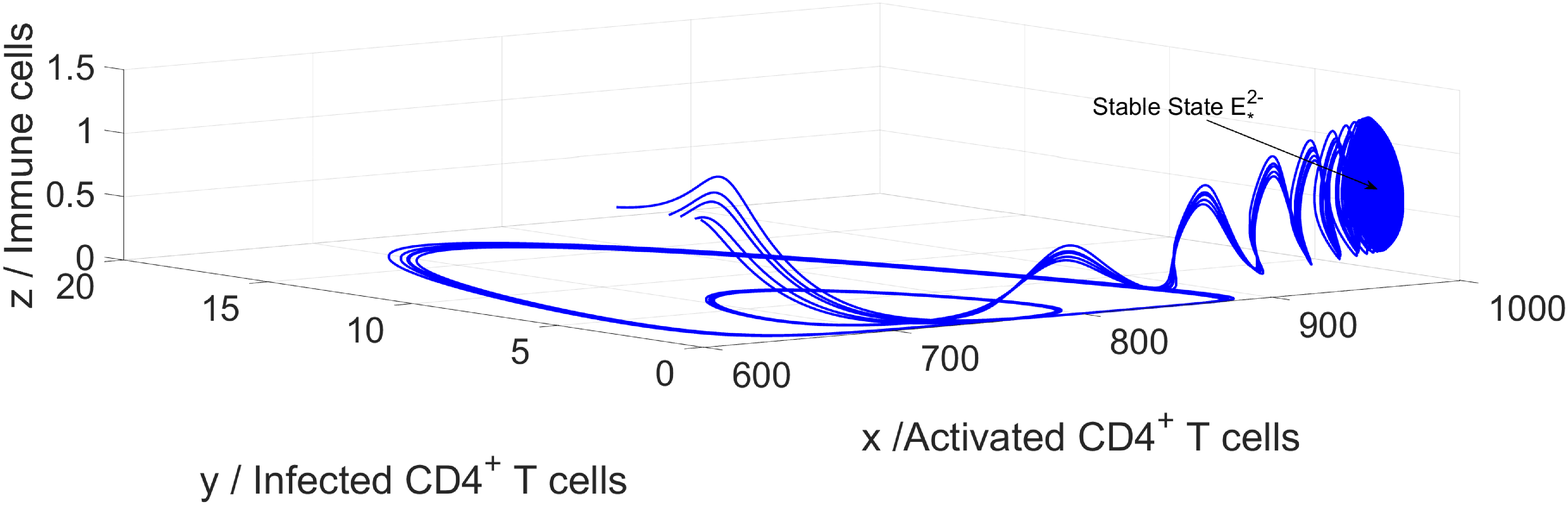}}}
 \caption{
\footnotesize System (1.3) has only the positive equilibrium
$E_{-}^{2*}$ is stable. Parameter $c=0.65~~\mbox{day}^{-1}$ and
other parameter values are shown in (3.1). We choose different
initial values. }\label{F51}
\end{center}
 \end{figure}

\section{2D-Viral infection system with monotonic immune response}

In this section, we discuss 2D viral infection system with monotonic
immune response.
  $$\left\{\begin
 {array}{l l}
\frac{dy}{dt}=\gamma y(1-\frac{y}{K})-ay-pyz=P_1,\\
\frac{dz}{dt}=f(y)z-bz=Q_1,
  \end {array}
  \right. \eqno(4.1) $$
where $f(y)$ is a monotonic function of $y$ and satisfies (1.2).

 System (4.1) always has an uninfected steady equilibrium
$E^{(3)}_{0}=(0,0)$, and if ${R}^{(2)}_0>1$, system (4.1) also has
an immune-free equilibrium $E^{(3)}_{1}=(y^{(3)}_{1}, 0)$; If
$\mathcal {R}_{*}^{(2)}>1$ system (4.1) has three equilibria
$E^{(3)}_{0}$, $E^{(3)}_{1}$ and $E_{*}^{(3)}=(y_{*}^{(3)},
z_{*}^{(3)})$, where
$$\begin
 {array}{ll}
 \displaystyle
  y^{(3)}_{1}=\frac{aK}{\gamma}({R}^{(3)}_{0}-1),\\\vspace{2\jot}
 y^{(3)}_{*}=f^{-1}(b),\\\vspace{2\jot} \displaystyle
z^{(3)}_{*}=\frac{a}{p}(\mathcal {R}_{*}^{(2)}-1).\end {array}
$$

We give a threshold
$$\mathcal
{R}^{(3)}_0=\frac{\gamma}{a},$$ and the basic immune reproductive
number is
$$\mathcal
{R}_{*}^{(2)}=\frac{\gamma}{a}(1-\frac{y^{(3)}_{*}}{K}).$$ This
ratio describes the average number of newly infected cells generated
form on infected cell at the beginning of the infectious process.

Let $\tilde{E}$ be any arbitrary equilibrium of system (4.1). The
Jacobian matrix associated with the system is
$$J_{3}=\left[
\begin{array}{cccc}
\gamma-a-\frac{2\gamma}{k}\tilde{y}-p\tilde{z} &-p\tilde{y}  \\
 f^{'}(\tilde{y})\tilde{z} & f(\tilde{y})-b \\
\end{array}
\right].$$ The characteristic equation of the linearized system of
(4.1) at $\tilde{E}$ is given by $\left|\lambda I-J_{3} \right|=0.$

\noindent{\bf Lemma 4.1}  $\mathcal {R}_{*}^{(2)}<1\Leftrightarrow
y^{(3)}_{1}<y_{*}^{(3)}$.

\noindent{\bf Proof.}
$$\begin
{array}{lll}
R_{*}^{(2)}<1&\Leftrightarrow& \frac{\gamma}{a}(1-\frac{y_{*}^{(3)}}{K})<1,  \\
&\Leftrightarrow & \frac{Ka}{\gamma}({R}^{(3)}_{0}-1)<  y_{*}^{(3)}, \\
&\Leftrightarrow& y^{(3)}_{1}< y_{*}^{(3)}.
\end {array}$$
\qed

\noindent{\bf Lemma 4.2} System (4.1) has no limit cycles in the
interior of the first quadrant.

\noindent{\bf Proof.} Consider the Dulac function
$$D_1=\frac{1}{yz}.$$
We can get

$$\begin {array}{lll}
\begin {split}
&\displaystyle \frac{\partial(D_1P_1)}{\partial y}+\frac{\partial(D_1Q_1)}{\partial z}\\
&=\frac{\partial[\frac{1}{yz}(\gamma y(1-\frac{y}{K})-ay-pyz)]}{\partial y}+\frac{\partial[\frac{1}{yz}(f(y)z-bz)]}{\partial z} \vspace{0.2cm}\\
&=\displaystyle
\frac{\partial(\frac{\gamma}{z}-\frac{\gamma y}{Kz}-\frac{a}{z}-p)}{\partial y}+\frac{\partial(\frac{f(y)}{y}-\frac{b}{y})}{\partial z} \vspace{0.2cm}\\
&= -\frac{\gamma}{Kz}\leq0.
\end{split}
\end {array}$$
By $Bendixson-Dulac$ discriminant method, we know system (4.1) has
no limit cycles. \qed

 \noindent{\bf Theorem 4.1} ~~If $\mathcal
{R}^{(3)}_0<1$, then the uninfected equilibrium $E^{(3)}_{0}$ of
system (4.1) is not only locally asymptotically stable, but also
global asymptotically stable. If $\mathcal {R}^{(3)}_0>1$. then  the
uninfected equilibrium $E^{(3)}_{0}$ of system (4.1) is unstable.

\noindent{\bf Proof.}  The characteristic equation of the linearized
system of system (4.1) at $E^{(3)}_{0}$  is
$$(\lambda+a-\gamma)(\lambda+b)=0.$$

Obviously, the characteristic roots  $-b$ and $a(\mathcal
{R}^{(3)}_{0} -1)$ are negative for $\mathcal {R}^{(3)}_0<1$. Hence
$E^{(3)}_{0}$ is locally asymptotically stable. If $\mathcal
{R}^{(3)}_0>1$, then $a(\mathcal {R}^{(3)}_0-1)>0$, thus, the
uninfected equilibrium $E^{(3)}_{0}$ of system (4.1) is unstable. By
Lemma 4.2, the uninfected equilibrium $E^{(3)}_{0}$ is global
asymptotically stable. Theorem 4.1 is proved. \qed

\noindent{\bf Theorem 4.2} ~~If $\mathcal {R}^{(3)}_0>1>\mathcal
{R}^{(2)}_{*}$, then the immune-free equilibrium $E^{(3)}_{1}$ of
system (4.1) is not only locally asymptotically stable, but also
global asymptotically stable. $E^{(3)}_{1}$ is unstable for
$\mathcal {R}^{(2)}_{*}>1$.

\noindent{\bf Proof.} The characteristic equation  of the linearized
system of (4.1) at $E^{(3)}_{1}$ is given by
$$(\lambda+\frac{\gamma}{k}y^{(3)}_{1})[\lambda-(f(y^{(3)}_{1})-b)]=0.$$
By Lemma 4.1 and $f^{'}(y)>0$ for $[0, +\infty)$ and
$f(y^{(3)}_{*})=b$, we deduce the eigenvalue
$\lambda=f(y^{(3)}_{1})-b<0$ for $\mathcal {R}^{(3)}_0>1>\mathcal
{R}^{(2)}_*$, and $\lambda=f(y^{(3)}_{1})-b>0$ for ${R}^{(2)}_*>1$.
Thus, the immune-free equilibrium $E^{(3)}_{1}$ of system (4.1) is
locally asymptotically stable for  $\mathcal {R}^{(3)}_0>1>\mathcal
{R}^{(2)}_*$ and is unstable for $\mathcal {R}^{(2)}_*>1$. By Lemma
4.2, the immune-free equilibrium $E^{(3)}_{1}$ is global
asymptotically stable. Theorem 4.2 is proved. \qed

\noindent{\bf Theorem 4.3} ~~If $\mathcal {R}^{(2)}_*>1$, then the
positive equilibrium $E^{(3)}_*$ of system (4.1) is not only locally
asymptotically stable, but also global asymptotically stable.

\noindent{\bf Proof.} The characteristic equation  of the linearized
system of (4.1) at $E^{(3)}_*$ is given by
$$\lambda^2+a^{(3)}_1\lambda+a^{(3)}_2=0,$$
where
$$\begin
{array}{lll}
a^{(3)}_1&=&\frac{\gamma}{k}y^{(3)}_*+b-f(y^{(3)}_*),\\
a^{(3)}_2&=&\frac{\gamma}{k}y^{(3)}_*[b-f(y^{(3)}_*)]+py^{(3)}_*z^{(3)}_*f^{'}(y^{(3)}_*).
\end {array}$$
By Lemma 4.1 and $f^{'}(y)>0$ for $[0, +\infty)$ and
$f(y^{(3)}_*)=b$, we know $a^{(3)}_1>0$ and $a^{(3)}_2>0$. By
Routh-Hurartz Criterion, we know the positive equilibrium
$E^{(3)}_*$ of system (4.1) is locally asymptotically stable for
$\mathcal {R}^{(2)}_*>1$. By Lemma 4.2, the positive equilibrium
$E^{(3)}_*$ is global asymptotically stable. Theorem 4.3 is proved.
\qed

By Theorem 4.1$\sim$4.3, we can get following result:

\noindent{\bf Remark 4.1 } Viral infection system with monotonic
immune response has no bistability appear.

\section{2D-Viral infection system with nonmonotonic immune response}

In this section, we will discuss the 2D-viral infection system with
Monod-Haldane function, which is a system with nonmonotonic immune
response.

  $$\left\{\begin
 {array}{l l}
\frac{dy}{dt}=\gamma y(1-\frac{y}{K})-ay-pyz=P_2,\\
\frac{dz}{dt}=\frac{cyz}{\alpha+\gamma y+y^2}-bz=Q_2,
  \end {array}
  \right. \eqno(5.1) $$

We always assume $K>\sqrt{\alpha}$. The threshold
${R}^{(4)}_0=\frac{\gamma}{a}$, which is equivalent to
${R}^{(3)}_0$.

(i) System (5.1) always has an uninfected steady equilibrium
$E^{(4)}_{0}=(0,0)$, and if $R^{(4)}_0>1$, system (5.1) also has an
immune-free equilibrium $E^{(4)}_{1}=(y^{(4)}_{1}, 0)$, where
$y^{(4)}_{1}=\frac{Ka}{\gamma}({R}^{(4)}_{0}-1)$.

Solving equation $\frac{cy}{  \alpha+\gamma y+y^2}-b=0$, one get two
positive roots, $c_{1}=\gamma b-2b\sqrt{\alpha}$ and $c_{2}=\gamma
b+2b\sqrt{\alpha}$ , then  the existence conditions of positive
equilibria as following:

(ii) If  $\mathcal {R}_{*}^{2-}>1$  and  $c>c_{2},$ system (5.1) has
an immune equilibrium $E_{*}^{4-}=(y_{*}^{4-}, z_{*}^{4-})$; If
$\mathcal {R}_{*}^{2+}>1$ and $c>c_{2},$ system (1.3) also has an
immune equilibrium $E_{*}^{4+}=(y_{*}^{4+}, z_{*}^{4+}).$ Here
$\mathcal {R}^{2\pm}_{*}=\frac{\gamma}{a}(1-\frac{y^{4\pm}_{*}}{K}),
y^{4\pm}_{*}=\frac{-B\pm\sqrt{B^{2}-4\alpha b^2}}{2b},
z^{4\pm}_{*}=\frac{a}{p}({R}^{2\pm}_{*}-1),  B= \gamma b-c.$

We denote post-treatment control threshold $P_{II}$ (see e.g. Refs
\cite{[19]})
$$c_{2}=\gamma b+2b\sqrt{\alpha}.$$
Which is equivalent to post-treatment control threshold $P_{I}$ .

Denote
 $$c^{*}_2=\gamma b+\frac{2baK(\mathcal{R}^{(4)}_0-1)}{\gamma},$$
$$c^{**}_2=\gamma b+\frac{baK(\mathcal {R}^{(4)}_0-1)}{\gamma}+\frac{b\alpha \gamma}{aK( {R}^{(4)}_0-1)},$$
We call $c^{**}_2$ the elite control threshold $E_{II}$, \cite{[19]}
which means the virus will be under control when the immune
intensity $c$ is larger than $c^{**}_2$.

Denote another threshold
 $$\mathcal{R}^{(2)}_{c}=1+\frac{\sqrt{\alpha}}{K-\sqrt{\alpha}}.$$

For the positive parameters in model (5.1), we have the following
lemmas.

\noindent{\bf Lemma 5.1 } $\mathcal {R}^{(4)}_0>\mathcal
{R}^{(2)}_{c}>1\Leftrightarrow c^{*}_2>c^{**}_2.$

\noindent{\bf Proof.}
$$\begin
{array}{lll} c^{*}_2>c^{**}_2&\Leftrightarrow&
\frac{baK({R}^{(4)}_0-1)}{\gamma}>\frac{b\alpha \gamma}{aK(\mathcal {R}^{(4)}_0-1)},  \\
&\Leftrightarrow&\mathcal {R}^{(4)}_0>\mathcal {R}^{(2)}_{c}.
\end {array}$$
\qed

\noindent{\bf Lemma 5.2 } (i) $\mathcal {R}^{(4)}_0>\mathcal
{R}^{(2)}_{c}>1\Leftrightarrow c^{*}_2>c_{2}$; (ii) $1<\mathcal
{R}^{(4)}_0<\mathcal {R}^{(2)}_{c}\Leftrightarrow c^{*}_2<c_{2}.$

\noindent{\bf Proof.}
$$\begin
{array}{lll} c^{*}_2>c_{2}&\Leftrightarrow&\frac{baK(\mathcal{R}^{(4)}_0-1)}{\gamma}>b\sqrt{\alpha},  \\
&\Leftrightarrow&\mathcal {R}^{(4)}_0>R^{(2)}_{c}.
\end {array}$$
$$\begin
{array}{lll} c^{*}_2<c_{2}&\Leftrightarrow&\frac{baK(\mathcal{R}^{(4)}_0-1)}{\gamma}<b\sqrt{\alpha},  \\
&\Leftrightarrow&\mathcal {R}^{(4)}_0<R^{(2)}_{c}.
\end {array}$$
\qed

\noindent{\bf Lemma 5.3 } (i) Assume $1<\mathcal
{R}^{(4)}_0<\mathcal {R}^{(2)}_{c}.$ If $\mathcal {R}_{*}^{2-}>1$,
then $c>c^{**}_2$; (ii) Assume $\mathcal {R}^{(4)}_0>\mathcal
{R}^{(2)}_{c}>1.$  If $\mathcal {R}_{*}^{2-}>1$, then $c>c_{2}$.

\noindent{\bf Proof.}
$$\begin
{array}{lll} \mathcal
{R}_{*}^{2-}>1&\Leftrightarrow& \frac{\gamma}{a}(1-\frac{y_{*}^{4-}}{K})>1,  \\
&\Leftrightarrow&\sqrt{(\gamma b-c)^2-4\alpha b^2}>c-c^{*}_2.
\end {array}$$
If $c<c^{*}_2$ and one of conditions  $c<c_{1}$ or $c>c_{2}$ is
correct, then $\mathcal {R}_{*}^{2-}$ is always larger than one. If
$c>c^{*}_2$, solving $\sqrt{(\gamma b-c)^2-4\alpha b^2}>c-c^{*}_2$,
we have $c>c^{**}_2.$ Thus,

(i) If $1<\mathcal {R}^{(4)}_0<\mathcal {R}^{(2)}_{c}$, then
$c^{*}_2<c_{2}$. From $\mathcal {R}_{*}^{2-}>1$, we have
$c>c^{**}_2.$

(ii) If $\mathcal {R}^{(4)}_{0}>\mathcal {R}^{(2)}_{c}>1$, then
$c^{*}_2>c_{2}$. From $\mathcal {R}_{*}^{2-}>1$, we have  $c>c_{2}.$
\qed

\noindent{\bf Lemma 5.4 } (i) If $1<\mathcal {R}^{(4)}_0<\mathcal
{R}^{(2)}_{c},$ then $\mathcal {R}_{*}^{2+}>1$ has no  solution;
(ii) Assume $\mathcal {R}^{(4)}_0>\mathcal {R}^{(2)}_{c}>1$. If
$\mathcal {R}_{*}^{2+}>1$, then $c_{2}<c<c^{**}_2$.

 \noindent{\bf Proof.}
$$\begin
{array}{lll} \mathcal {R}_{*}^{2+}>1
&\Leftrightarrow& \frac{\gamma}{a}(1-\frac{y_{*}^{4+}}{K})>1,\\
&\Leftrightarrow&c^{*}-c>\sqrt{(\gamma b-c)^2-4\alpha b^2}.
\end {array}$$

(i) If $1<{R}^{(4)}_0<R^{(2)}_{c},$ then $c^{*}_2<c_{2}$. Thus
$\mathcal {R}_{*}^{2+}>1$ has no  solution. (ii) If $\mathcal
{R}^{(4)}_0>\mathcal {R}^{(2)}_{c}>1$, then $c^{*}_2>c_{2}$. Solving
$\mathcal {R}_{*}^{2+}>1$, we have $c_{2}<c<c^{**}_2$. \qed

By Lemma 5.1 $\sim$ Lemma 5.4  and summing up the above analysis we
obtain the existing results of equilibria of system (5.1).

\noindent{\bf Theorem 5.1} (i) System (5.1) always exists an
uninfected equilibrium $E^{(4)}_{0}=(0,0);$

(ii) If  $\mathcal {R}^{(4)}_0>1$, system (5.1) also has an
immune-free equilibrium $E^{(4)}_{1}=(y^{(4)}_1,0),$ where
$y^{(4)}_1=\frac{aK}{\gamma}(R^{(4)}_0-1)$;

(iii) If $1<\mathcal {R}^{(4)}_0<\mathcal {R}^{(2)}_{c}$ and
$c>c^{**}_2,$ system (5.1) also has  one positive equilibrium
$E_{*}^{4-};$

(iv) If $\mathcal {R}^{(4)}_0>\mathcal {R}^{(2)}_{c}>1$ and
 $c_{2}<c<c^{**}_2$, system (5.1)  has  two positive equilibria
$E_{*}^{4+}$ and  $E_{*}^{4-}$. While  $\mathcal
{R}^{(4)}_0>\mathcal {R}^{(2)}_{c}$ and $c>c^{**}_2$, system (5.1)
only  has  one positive equilibrium $E_{*}^{4-}$;

The summary results of the existence for positive equilibria can be
seen in Table 5 and Table 6.

\subsection{Stability analysis}
Let $\tilde{E}$ be any arbitrary equilibrium of system (5.1). The
Jacobian matrix associated with the system is

$$J_{4}=\left[
\begin{array}{cccc}
 \gamma-a-\frac{2\gamma}{K}y-pz  &  -py \\
\frac{(\alpha-y^2)cz}{(\alpha+\gamma y+y^2)^2}  & \frac{cy}{\alpha+\gamma y+y^2}-b \\
\end{array}
\right].$$ The characteristic equation of the linearized system of
(5.1) at $\tilde{E}$ is given by $\left|\lambda I-J_{4} \right|=0.$

\noindent{\bf Lemma 5.5} System (5.1) has no limit cycles in the
interior of the first quadrant.

\noindent{\bf Proof.} Consider the Dulac function
$$D_2=\frac{1}{yz}.$$
We can get

$$\begin {array}{lll}
\begin {split}
&\displaystyle
\frac{\partial(D_2P_2)}{\partial y}+\frac{\partial(D_2Q_2)}{\partial z}\\
&=\frac{\partial[\frac{1}{yz}(\gamma y(1-\frac{y}{K})-ay-pyz)]}{\partial y}+\frac{\partial[\frac{1}{yz}(\frac{cyz}{\alpha+\gamma y+y^2}-bz)]}{\partial z}\\
&=\displaystyle \frac{\partial(\frac{\gamma}{z}-\frac{\gamma y}{Kz}-\frac{a}{z}-p)}{\partial y}+\frac{\partial(\frac{c}{\alpha+\gamma y+y^2}-\frac{b}{y})}{\partial z} \vspace{0.2cm}\\
&= -\frac{\gamma}{Kz}\leq0.
\end{split}
\end {array}$$

By Bendixson-Dulac discriminant method, we know system (5.1) has no
limit cycles.\ \qed

\noindent{\bf Theorem 5.2} ~~If $\mathcal {R}^{(4)}_0<1$, then the
uninfected equilibrium $E^{(4)}_{0}$ of system (5.1) is not only
locally asymptotically stable, but also global asymptotically
stable. If $\mathcal {R}^{(4)}_0>1$. then the uninfected equilibrium
$E^{(4)}_{0}$ of system (5.1) is unstable.

\noindent{\bf Proof.}  The characteristic equation of the linearized
system of system (5.1) at $E^{(4)}_{0}$  is
$$(\lambda+a-\gamma)(\lambda+b)=0.$$
Obviously, the characteristic roots  $-b$ and $a(\mathcal
{R}^{(4)}_{0} -1)$ are negative for $\mathcal {R}^{(4)}_0<1$. Hence
$E^{(4)}_{0}$ is locally asymptotically stable.  If $\mathcal
{R}^{(4)}_0>1$, then $a(\mathcal {R}^{(4)}_0-1)>0$, thus, the
uninfected equilibrium $E^{(4)}_{0}$ of system (5.1) is unstable. By
Lemma 5.5, the uninfected equilibrium $E^{(4)}_{0}$ is global
asymptotically stable. Theorem 5.2 is proved. \qed

\noindent{\bf Theorem 5.3} ~~If $\mathcal {R}^{(4)}_0>1$ and
$0<c<c^{**}_2$, then the immune-free equilibrium $E^{(4)}_{1}$ of
system (5.1) is not only locally asymptotically stable, but also
global asymptotically stable.

\noindent{\bf Proof.} The characteristic equation  of the linearized
system of (5.1) at $E^{(4)}_1$ is given by
$$[\lambda-(\gamma-a-\frac{2\gamma}{K}y^{(4)}_1)][\lambda-(\frac{cy^{(4)}_1}{\alpha+\gamma y^{(4)}_1+{(y^{(4)}_1})^2}-b)]=0,$$ we get two eigenvalues $\lambda_1=\gamma-a-\frac{2\gamma}{K}y^{(4)}_1=a(1-\frac{\gamma}{a})<0$ for $\mathcal
{R}^{(4)}_{0}>1$, and $\lambda_2=\frac{cy^{(4)}_1}{\alpha+\gamma
y^{(4)}_1+{(y^{(4)}_1})^2}-b>0$ for $0<c<c^{**}_2$. Thus, the
immune-free equilibrium $E^{(4)}_{1}$ of system (5.1) is locally
asymptotically stable for $\mathcal {R}^{(4)}_{0}>1$ and
$0<c<c^{**}_2$. By Lemma 5.5, the immune-free equilibrium
$E^{(4)}_{1}$ is global asymptotically stable. Theorem 5.3 is
proved. \qed

 \noindent{\bf Theorem 5.4}
(i) If \hspace{0.1cm} ($\mathbf{A.1}$) \hspace{0.2cm} $1<\mathcal
{R}^{(4)}_0<\mathcal {R}^{(2)}_{c}$ and $c>c^{**}_2$, or

\hspace{2.3cm} ($\mathbf{A.2}$) \hspace{0.2cm}    $\mathcal
{R}^{(4)}_0>\mathcal
{R}^{(2)}_{c}$ and $c>c_{2}$, \\
system (5.1) has an immune equilibrium $E_{*}^{4-},$ which is not
only asymptotically stable, but also global asymptotically stable.

(ii) If  $\mathcal {R}^{(4)}_0>\mathcal {R}^{(2)}_{c}$ and
 $c_{2}<c<c^{**}_2$, system (5.1) also has an immune
 equilibrium $E_{*}^{4+},$ which is an unstable saddle.

\noindent{\bf Proof.} Denote $E_{*}^{(4)}=(y_{*}^{(4)},
z_{*}^{(4)})$ as an arbitrary positive equilibrium of system (5.1).
The characteristic equation of the linearized system of (5.1) at the
arbitrary positive equilibrium $E_{*}^{(4)}$ is given by
$$\lambda^2+b^{(4)}_1\lambda+b^{(4)}_2=0,$$ where
 $$\begin
{array}{lll}
b^{(4)}_1&=&\frac{\gamma}{K}y_{*}^{(4)}>0,\\
b^{(4)}_2&=&\frac{\alpha-(y_{*}^{(4)})^2}{(\alpha+\gamma
y_{*}^{(4)}+(y_{*}^{(4)})^2)^2}pcy_{*}^{(4)}z_{*}^{(4)}.
\end {array}$$

For equilibrium $E_{*}^{4-},$
$$\begin
{array}{lll} \alpha-(y_{*}^{4-})^2>0
&\Leftrightarrow&\frac{-B-\sqrt{B^2-4\alpha b^2}}{2b}<\sqrt{\alpha} ,\\
&\Leftrightarrow&c>c_2.
\end {array}$$
If $c>c_2$, we can get $b^{(4)}_2>0$, by Routh-Hurartz Criterion, we
know in this case the positive equilibrium $E_{*}^{4-}$ is a stable
node.

For equilibrium $E_{*}^{4+},$
$$\begin
{array}{lll} \alpha-(y_{*}^{4+})^2<0
&\Leftrightarrow&\frac{-B+\sqrt{B^2-4\alpha b^2}}{2b}>\sqrt{\alpha} ,\\
&\Leftrightarrow& \sqrt{B^2-4\alpha B^2}>B+2b\sqrt{\alpha}.
\end {array}$$
If $c_2<c<c^{**}_2$, then $b^{(4)}_2<0$, so the immune equilibrium
$E_{*}^{4+}$ is an unstable saddle. By Lemma 5.5, the immune
equilibrium $E_{*}^{4-},$ is global asymptotically stable. Theorem
5.4 is proved. \qed

\subsection{Saddle-node Bifurcation }

If $\mathcal {R}^{(4)}_{0}>\mathcal {R}^{(2)}_{c}>1$ and
$c^2-2\gamma bc+\gamma^2b^2-4\alpha b^2=0$, the immune equilibrium
$E_{*}^{4+}$ and  $E_{*}^{4-}$ coincide with each other. Then system
has the unique interior equilibrium $E_{*}=(y_{*},
z_{*})=(\sqrt{\alpha}, \frac{a}{pR_c}(R_0-R_c))$. The emergence and
disappearance of the equilibrium is due to the occurrence of
saddle-node bifurcation when $c$ crosses the bifurcation value
$c^{[sn]}$, where $c^{[sn]}=\gamma b+2b\sqrt{\alpha}$.

\noindent{\bf Theorem 5.5 } If $\mathcal {R}^{(4)}_{0}>\mathcal
{R}^{(2)}_{c}>1$ and $c=c^{[sn]}$, system (5.1) will undergoes a
saddle-node bifurcation, $c$ as the bifurcation parameter is given
by $c=c^{[sn]}=\gamma b+2b\sqrt{\alpha}$.

\noindent{\bf Proof.} We use Sotomayor's theorem
\cite{[27],[28],[29]} to prove system (5.1) undergoes a saddle-node
bifurcation at $c=c^{[sn]}$. It's easy to prove $Det[J_{E_{*}}]=0$,
so one of the eigenvalue of the Jacobian at the saddle-node
equilibrium is zero, where $J=J_4$.

Let $V=(V_1, V_2)^\mathrm{T}$ and $W =(W_1, W_2)^\mathrm{T}$
represent the eigenvectors of $J_{E_{*}}$ and $J_{E_{*}}^\mathrm{T}$
corresponding to the zero eigenvalue, respectively, then they are
given by $V =(1, -\frac{\gamma}{Kp})^\mathrm{T}$ and $W=(0,
1)^\mathrm{T}$. Let $F=(P_2, Q_2)$, we can get
 $$F_c(E_*;c^{[sn]})=\left [
\begin{array}{cccc}
0\\
\frac{yz}{\alpha+\gamma y+y^2}
\end{array}
\right ]_{(E_*;c^{[sn]})}= \left[ \begin{array}{cccc}
0 \\
\frac{\gamma(1-\frac{\sqrt{\alpha}}{K})-a}{p(2\sqrt{\alpha}+\gamma)}
\end{array}
\right ],
$$
and
$$\begin {array}{lll}
\begin {split}
\displaystyle & D^2F(E_*;c^{[sn]})(V,V)\\ &= \left [
\begin{array}{cccc}
0\\
\frac{-6c\alpha yz+2czy^3-2c\alpha\gamma z}{(\alpha+\gamma
y+y^2)^3}-\frac{2\gamma c(\alpha-y^2)}{pK(\alpha+\gamma y+y^2)^2}
\end{array}
\right ]_{(E_*;c^{[sn]})}\\
&=\left[
\begin{array}{cccc}
0 \\
\frac{-2\alpha z_*(\gamma
b+2b\sqrt{\alpha})(2\sqrt{\alpha}+\gamma)}{(\alpha+\gamma
y_*+y_*^2)^3}
\end{array}
\right ].
\end{split}
\end {array}$$

Therefore,
$$\begin {array}{lll}
\begin {split}
\Omega_1&=W^\mathrm{T}F_c(E_*,c^{[sn]})=\frac{\gamma(1-\frac{\sqrt{\alpha}}{K})-a}{p(2\sqrt{\alpha}+\gamma)}\neq0,   \vspace{0.2cm}\\
\Omega_2&=W^\mathrm{T}[D^2F(E_*;c^{[sn]})(V,V)]=\frac{-2\alpha
z_*(\gamma b+2b\sqrt{\alpha})(2\sqrt{\alpha}+\gamma)}{(\alpha+\gamma
y_*+y_*^2)^2}\neq0.
\end{split}
\end {array}$$

Therefore, from the Sotomayor' s theorem, \cite{[27],[28],[29]}
system (5.1) undergoes a saddle-node bifurcation at $E_{*}=(y_{*},
z_{*})$ when $c=c^{[sn]}$. Hence, we can conclude that when
parameter $c$ passes from one side from of $c=c^{[sn]}$ to the other
side, the number of interior equilibrium of system (5.1) changes
from zero to two.

\qed

\subsection{Transcritical Bifurcation }

From the stability analysis of system (5.1), the boundary
equilibrium $E_1^{(4)}$ looses its stability at $c=\gamma
b+\frac{baK(\mathcal {R}^{(4)}_0-1)}{\gamma}+\frac{b\alpha
\gamma}{aK(\mathcal {R}^{(4)}_0-1)}$ and one of the eigenvalue of
the Jacobian at $E_1^{(4)}$ is zero. Therefore, bifurcation may
occur at the boundary equilibrium $E_1^{(4)}$. In this section, we
select parameter $c$ as bifurcation parameter to study the existence
of a transcritical bifurcation.

\noindent{\bf Theorem 5.6 } If $R_0>1$ and $c=c^{[tc]}$, system
(5.1) will undergoes a transcritical bifurcation between $E_1^{(4)}$
and $E_*^{4-}$, $c$ as the bifurcation parameter is given by
$c=c^{[tc]}=\gamma b+\frac{baK({R}^{(4)}_0-1)}{\gamma}+\frac{b\alpha
\gamma}{aK({R}^{(4)}_0-1)}$.

\noindent{\bf Proof.} We use Sotomayor's theorem
\cite{[27],[28],[29]} to prove system (5.1) undergoes a
transcritical bifurcation. Obviously, one of the eigenvalue of the
Jacobian at $E_1^{(4)}$ is zero, if and only if $c=c^{[tc]}$.

Let $\nu =(\nu_1, \nu_2)^\mathrm{T}$ and $\omega=(\omega_1,
\omega_2)^\mathrm{T}$ denote the eigenvectors of $J_{E_1^{(4)}}$ and
$J_{E_1^{(4)}}^\mathrm{T}$ corresponding to the zero eigenvalue,
respectively, we can get $\nu =(1, -\frac{\gamma}{Kp})^\mathrm{T}$
and $\omega=(0, 1)^\mathrm{T}$, Besides,
$$F_c(E_1^{(4)};c^{[tc]})=\left [
\begin{array}{cccc}
0\\
\frac{yz}{\alpha+\gamma y+y^2}
\end{array}
\right ]_{(E_1^{(4)};c^{[tc]})}= \left[ \begin{array}{cccc}
0 \\
0
\end{array}
\right ].
$$
$$\begin {array}{lll}
\begin {split}
\displaystyle & DF_c(E_1^{(4)};c^{[tc]})\nu\\ &= \left [
\begin{array}{cccc}
0\\
\frac{\alpha z-zy^2}{(\alpha+\gamma y+y^2)^2}-\frac{\gamma
y}{Kp(\alpha+\gamma y+y^2)}
\end{array}
\right ]_{(E_1^{(4)};c^{[tc]})}\\
&=\left[
\begin{array}{cccc}
0 \\
-\frac{\gamma y_1^{(4)}}{Kp(\alpha+\gamma y_1^{(4)}+{y_1^{(4)}}^2)}
\end{array}
\right ].
\end{split}
\end {array}$$
$$\begin {array}{lll}
\begin {split}
\displaystyle & D^2F(E_1^{(4)};c^{[tc]})(\nu,\nu)\\ &= \left [
\begin{array}{cccc}
0\\
\frac{-6c\alpha yz+2czy^3-2c\alpha\gamma z}{(\alpha+\gamma
y+y^2)^3}-\frac{2\gamma c(\alpha-y^2)}{pK(\alpha+\gamma y+y^2)^2}
\end{array}
\right ]_{(E_1^{(4)};c^{[tc]})}\\
&=\left[
\begin{array}{cccc}
0 \\
\frac{-2\gamma(\gamma b+\frac{baK({R}^{(4)}_0-1)}{r}+\frac{b\alpha
r}{aK( {R}^{(4)}_0-1)})(\alpha-{y_1^{(4)}}^2)}{(\alpha+\gamma
y_1^{(4)}+{y_1^{(4)}}^2)^2}
\end{array}
\right ].
\end{split}
\end {array}$$
Therefore,
$$\begin {array}{lll}
\begin {split}
\Phi_1&=\omega^\mathrm{T}F_c(E_1^{(4)};c^{[tc]})=0,   \vspace{0.2cm}\\
\Phi_2&=\omega^\mathrm{T}[DF_c(E_1^{(4)};c^{[tc]})\eta]=-\frac{\gamma y_1^{(4)}}{Kp(\alpha+\gamma y_1^{(4)}+{y_1^{(4)}}^2)}\neq0\\
\Phi_3&=\omega^\mathrm{T}[D^2F(E_1^{(4)};c^{[tc]})(\nu,\nu)]=\frac{-2\gamma(\gamma
b+\frac{baK( {R}^{(4)}_0-1)}{\gamma}+\frac{b\alpha \gamma}{aK(
{R}^{(4)}_0-1)})(\alpha-{y_1^{(4)}}^2)}{(\alpha+\gamma
y_1^{(4)}+{y_1^{(4)}}^2)^2}\neq0.
\end{split}
\end {array}$$

Therefore, system (5.1) will undergoes a transcritical bifurcation
between $E_1^{(4)}$ and $E_*^{4-}$ at $c=c^{[tc]}$

\qed

 \noindent{\bf Remark 5.1 }   If $\mathcal
{R}^{(4)}_{0}>\mathcal {R}^{(4)}_{c}>1$ and
 $c_{2}<c<c^{**}_2$, system (5.1) has bistability appear. In other cases, system (5.1) has no bistability appear. Threshold
 $c_{2}$ is the post-treatment control threshold, $c^{**}_2$ is the elite control
 threshold. $(c_{2}, c^{**}_2)$ is the bistable interval.   \qed

To sum up, the stabilities of the equilibria and the behaviors of
system (5.1) can be shown in Table 7 and Table 8.

\subsection{Numerical simulations and discussion}
To verify our analysis results,  we carry out some numerical
simulations choosing some parameter values shown as in
\cite{[24],[25]}:
$$\begin
{array}{lll} &&\gamma=6 ~~\mbox{day}^{-1},\hspace{0.02cm}  K=6 ~~cells/\mu l, \hspace{0.02cm}  a=3~~\mbox{day}^{-1},\\
&& p=1~~\mbox{day}^{-1},\hspace{0.02cm}  \alpha=1~~cells/\mu l,\hspace{0.02cm}  \gamma=0.5~~cells/\mu l,\\
&& b=1~~\mbox{day}^{-1}.
\end {array}
   \eqno(5.1)$$

The parameters chose as same as in (5.1), the thresholds $\mathcal
{R}^{(4)}_{0}=2.0000$, $\mathcal {R}^{(2)}_{c}=1.2000,$
post-treatment control threshold $c_{2}=2.5000$ and elite control
threshold $c^{**}_2\approx3.5278$. In this case, $\mathcal
{R}^{(4)}_{0}>\mathcal {R}^{(2)}_{c}$ and $c_{2}<c^{**}_2$, then we
get a bistable interval $(2.5000, 3.8333)$(see Figure 5). When
$0<c<c_{2}$, the immune-free equilibrium $E^{(4)}_1$ is stable (see
Fig. 7); When $c_{2}<c<c^{**}_2$, the immune-free equilibrium
$E^{(4)}_1$ and the positive equilibrium $E_{*}^{4-}$ are stable
(see Fig. 6); When $c>c^{**}_2$, only the positive equilibrium
$E_{*}^{4-}$ is stable (see Figure 7).

\begin{figure}[!h]
\begin{center}
{\rotatebox{0}{\includegraphics[width=0.7 \textwidth,
height=50mm]{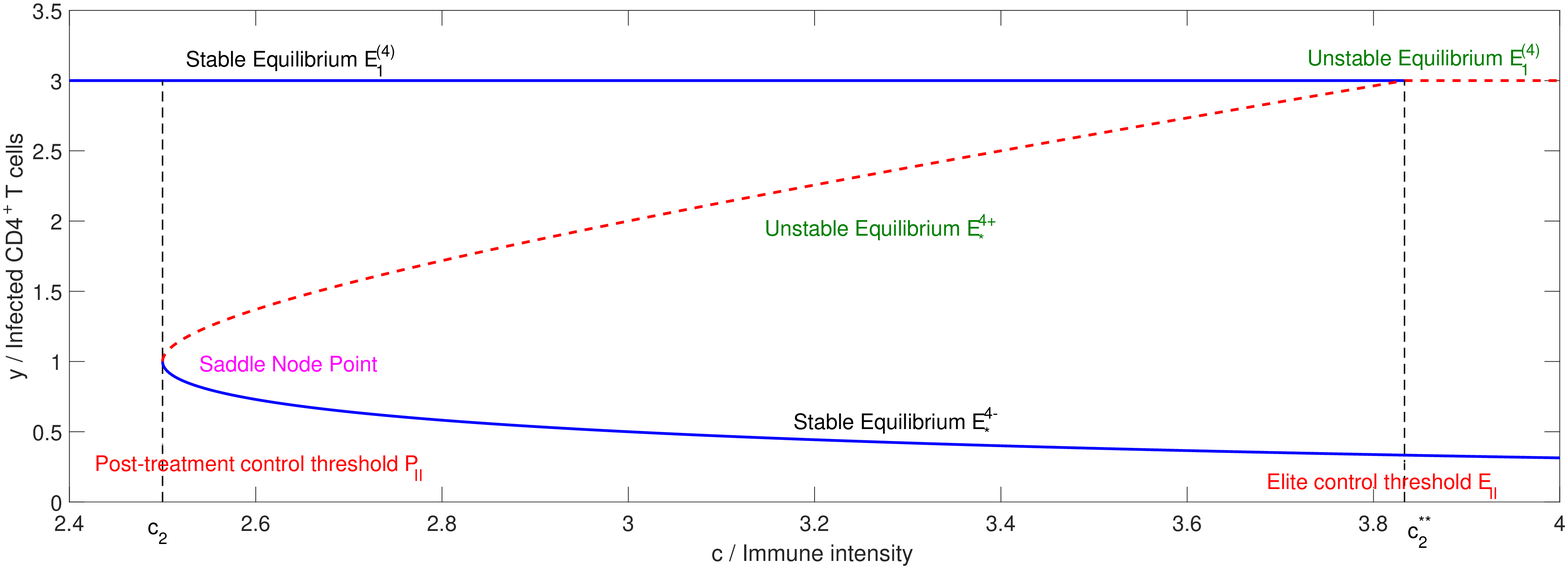}}}
 \caption{
\footnotesize Bistability and saddle-node bifurcation diagram of
system (1). The solid line is the stable virus and the dashed line
depends the unstable virus.  The post-treatment control threshold is
$c_{2}=2.5000$, the elite control threshold is
$c^{**}_2\approx3.5278$ and the bistable interval is $(2.5000,
3.5278).$ $c=3~~\mbox{day}^{-1}$ and other parameter values are
shown in (5.1). }\label{F51}
\end{center}
 \end{figure}

\begin{figure}[!h]
\begin{center}
{\rotatebox{0}{\includegraphics[width=0.28 \textwidth,
height=40mm]{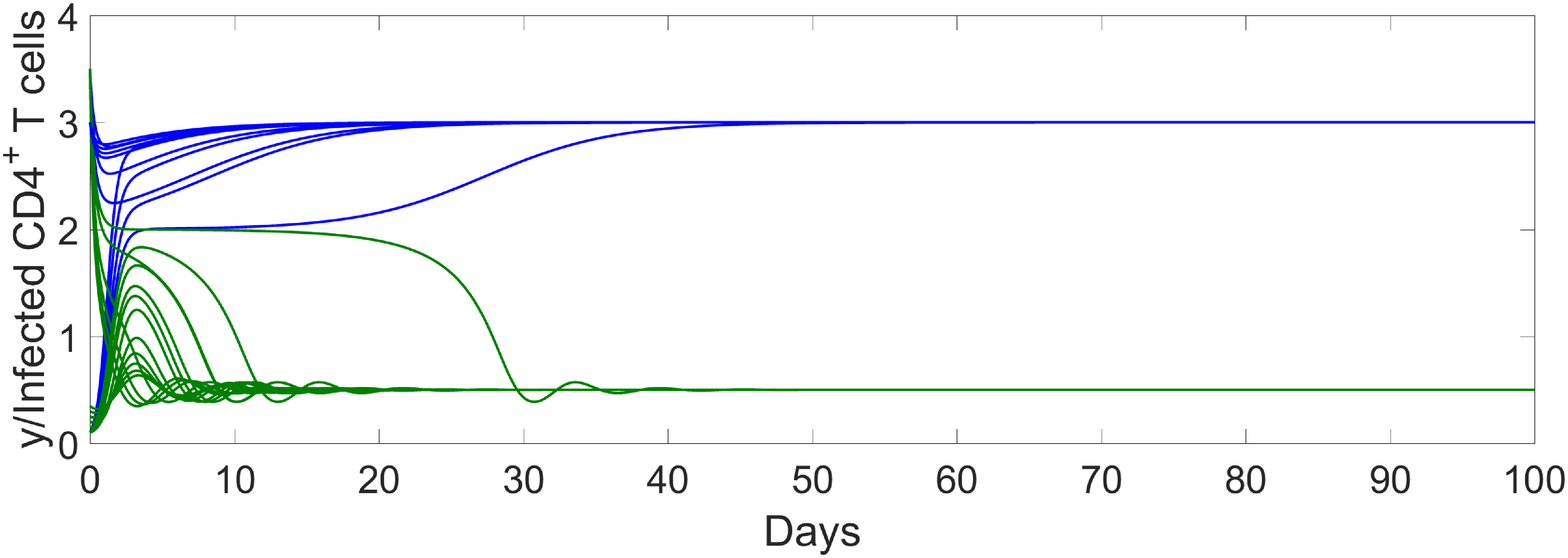}}} {\rotatebox{0}{\includegraphics[width=0.28
\textwidth, height=40mm]{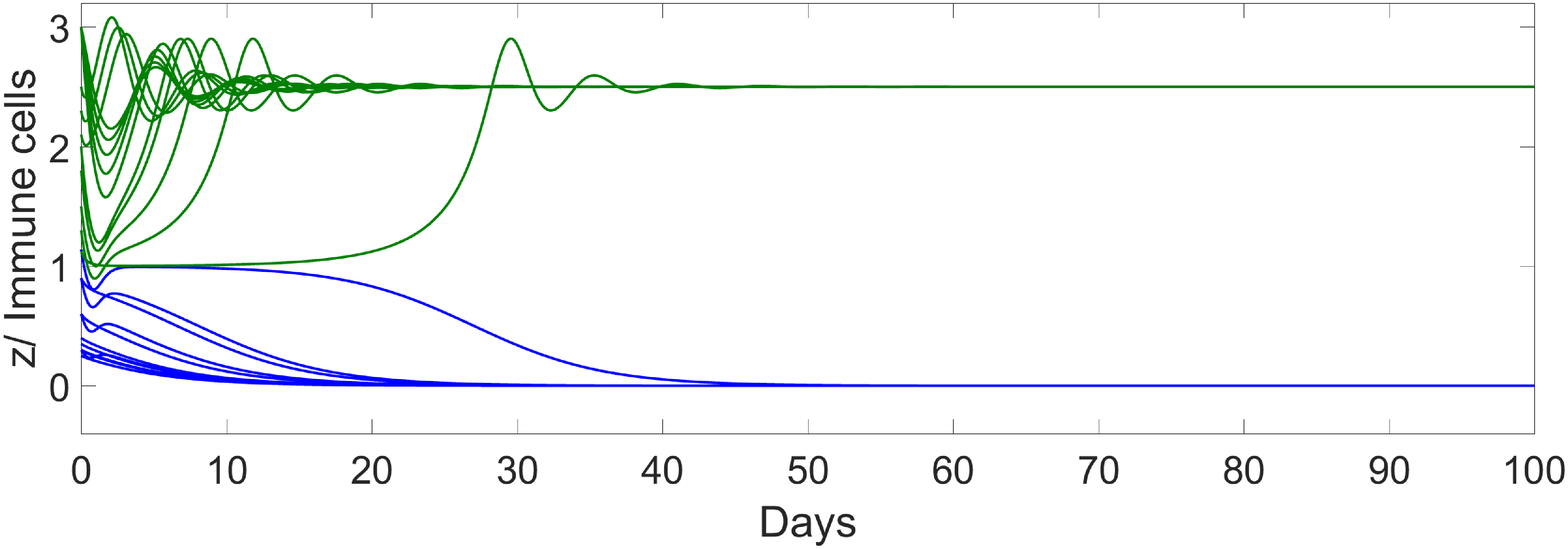}}}
{\rotatebox{0}{\includegraphics[width=0.34 \textwidth,
height=40mm]{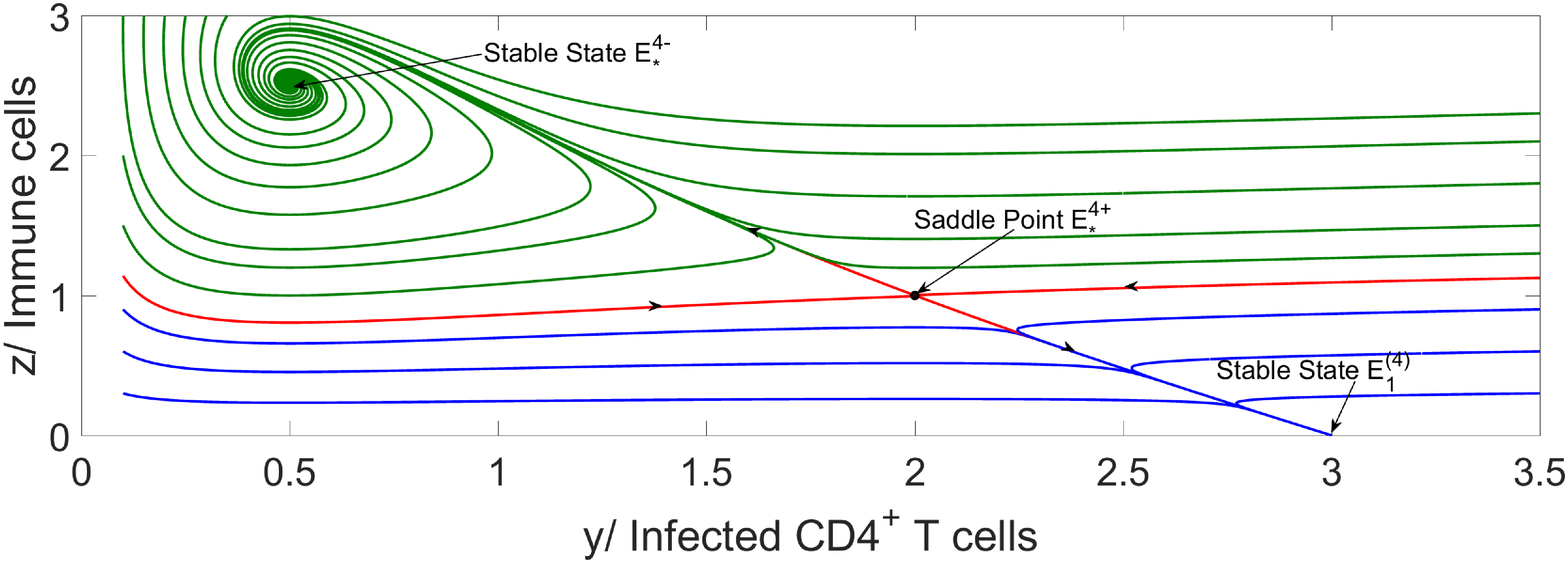}}}
 \caption{
\footnotesize System (1) has two different stable equilibria
$E^{(4)}_{1}$ and $E_{*}^{4-}$. Parameter $c=3~~\mbox{day}^{-1}$ and
other parameter values are shown in (5.1). We choose different
initial values. }\label{F51}
\end{center}
 \end{figure}
\begin{figure}[!h]
\begin{center}
{\rotatebox{0}{\includegraphics[width=0.48 \textwidth,
height=50mm]{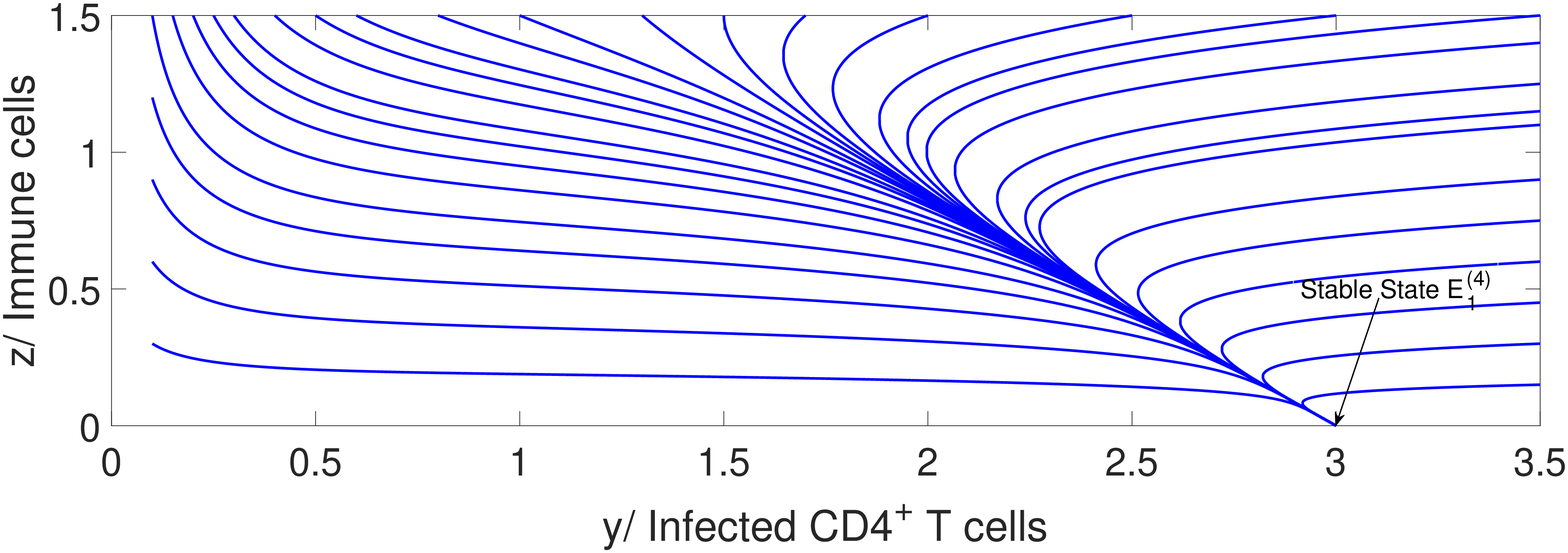}}} {\rotatebox{0}{\includegraphics[width=0.48
\textwidth, height=50mm]{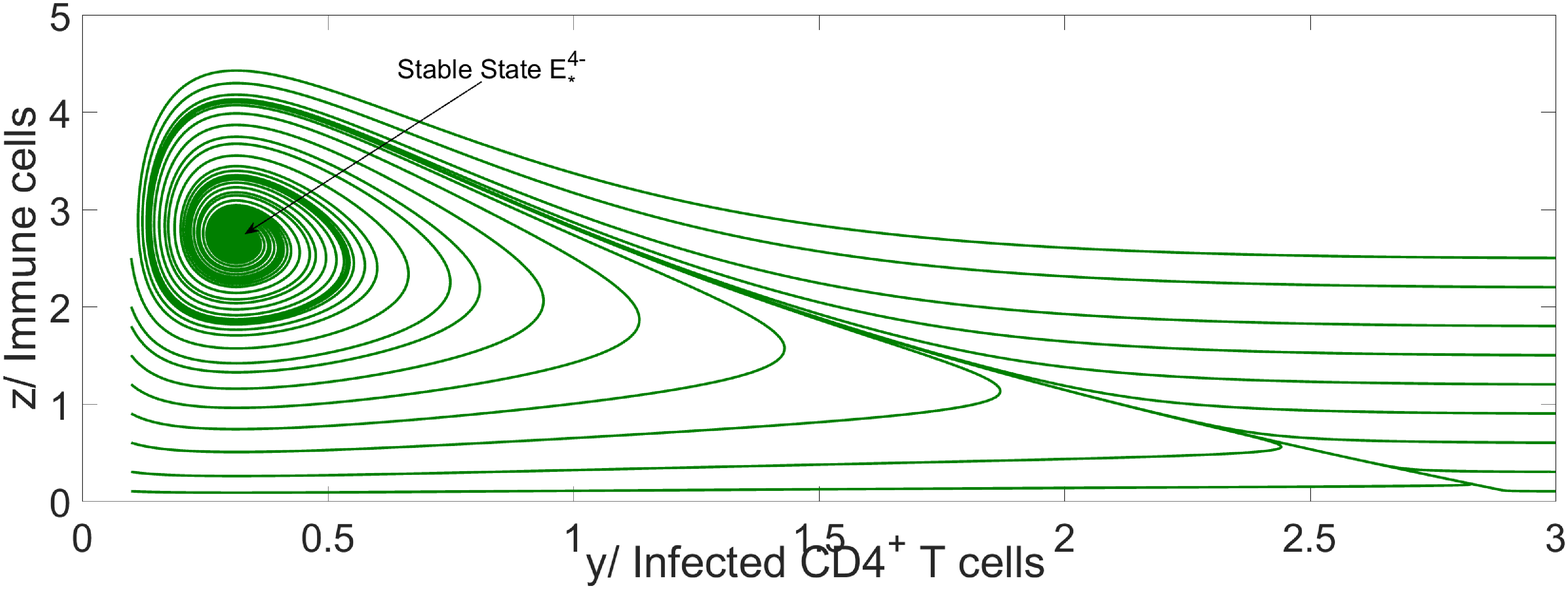}}}
 \caption{
\footnotesize (A) Choosing $c=2~~\mbox{day}^{-1}$, less than the
post-treatment control threshold $c_{2}=2.5000$, system (5.1) only
has a stable equilibrium $E^{(4)}_{1}$; (B) While choosing
$c=4~~\mbox{day}^{-1}$, larger than the elite control threshold
$c^{**}_2\approx3.5278$,  system (5.1) only has the stable
equilibria $E_{*}^{4-}$. Other parameter values are shown in (5.1).
}\label{F51}
\end{center}
 \end{figure}

\section{Discussion}
In this paper, we have considered the 2-dimensional, 3-dimensional
monotonic and nonmonotonic immune response in viral infection
system. For viral infection system with monotonic immune response,
by analyzing the existence and stability of the equilibria of the
viral infection system with monotonic immune response, we find that
the system with monotonic immune response has no bistability appear.
Beside, we discuss the viral infection system with nonmonotonic
immune response, and chose Monod-Haldane function as the
nonmonotonic immune response. For viral infection system with
nonmonotonic immune response, we find the system has bistability
appear under some conditions. Through calculations, we got two
important threshold. We call them post-treatment control threshold
and elite control threshold. Below the post-treatment control
threshold, the system has a stable immune-free steady state, which
means the viral will be rebound. Above the elite control threshold,
the system has a stable positive equilibrium, which indicates that
the virus will be under control. While between the two thresholds is
a bistable interval, the system can have bistability appear, which
imply that the patients either experience viral rebound after
treatment or achieve the post-treatment control. Select the rate of
immune cells stimulated by the viruses as a bifurcation parameter
for 2-dimensional and 3-dimensional nonmonotonic immune responses,
we prove the system exhibits saddle-node bifurcation and
transcritical bifurcation. The numerical simulations can help us
test the results of analysis and better understand the model.


\begin{thebibliography}{00}
\bibitem{[1]}C. Bartholdy, J.P. Christensen, D. Wodarz, A.R. Thomsen. Persistent virus infection despite chronic cytotoxic T-lymphocyte activation in Gamma interferon-deficient mice infected with lymphocytic chroriomeningitis virus, {\it J. Virol.} {\bf 74}(2000) 10304--10311.

\bibitem{[2]}W.M. Liu, Nonlinear oscillations in models of immune responses to persistent viruses, {\it Theor. Popul. Biol.} {\bf 52}(1997) 224--230.

\bibitem{[3]}M.A. Nowak, C.R.M. Bangham. Population dynamics of immune responses to persistent viruses, {\it Science} {\bf 272}(1996) 74--79.

\bibitem{[4]}D. Wodarz. Hepatitis C virus dynamics and pathology: The role of CTL and antibody responses, {\it J. Gen. Virol.} {\bf 84}(2003) 1743--1750.

\bibitem{[5]}D. Wodarz, J.P. Christensen, A.R. Thomsen. The importance of lytic and nonlytie immune responses in viral infections, {\it Trends Immunol.} {\bf 23}(2002) 194--200.

\bibitem{[6]}S. Bonhoeffer, R.M. May, G.M. Shaw, M.A. Nowak. Virus dynamics and drug therapy, {\it Proc. Natl. Acad. Sci.} {\bf 94}(1997) 6971--6976.

\bibitem{[7]}A.V.M. Herz, S. Bonhoeffer, R.M. Anderson, R.M. May, M.A. Nowak. Viral dynamics in vivo: Limitations on estimates of intracellular delay and virus decay, {\it Proc. Natl. Acad. Sci.} {\bf 93}(1996) 7247--7251.

 \bibitem{[8]} A. Korobeinikov. Global properties of basic virus dynamics models, {\it B. Math. Biol.} {\bf 66}(2004) 879--883.

\bibitem{[9]}P.D. Leenheer, H.L. Smith. Virus dynamics: A global analysis, {\it SIAM J. Appl. Math.} {\bf 63}(2003) 1313--1327.

\bibitem{[10]}M.A. Nowak, S. Bonhoeffer, A. M. Hill, R. Boehme, H. C. Thomas. Viral dynamics in hepatitis B virus infection, {\it Proc. Natl. Acad. Sci.} {\bf 93}(1996) 4398--4402.

\bibitem{[11]}K. Wang, Z. Qiu, G. Deng. Study on a population dynamic model of virus infection, {\it J. Sys. Sci. and Math. Scis.} {\bf 23}(2003) 433--443.

\bibitem{[12]}M.A. Nowak, C.R. M. Bangham. Population dynamics of immune
response to persistent viruses, {\it Science} {\bf 272} (1996)
74--79.

\bibitem{[13]}J.F. Andrews. A mathematical model for the continuous culture of microorganisms utilizing inhibitory substrates, {\it Biotechnol. Bioeng.} {\bf 10}(1968) 707--723.

\bibitem{[14]}W. Sokol, J.A. Howell. Kinetics of phenol oxidation by washed cells, {\it Biotechnol. Bioeng.} {\bf 23}(1980) 2039--2049.

\bibitem{[15]}S.L. Wang, F. Xu. L.B. Rong. Bistable analysis of an HIV model with immune response, {\it J. Bio. Syst.}{\bf 25}(4)(2017) 677--695.

\bibitem{[20]}M.A. Nowak, C.R.M. B angham. Population dynamics of immune response to persistent viruses. {\it Science} {\bf 272}(2)(1996).

\bibitem{[16]}F. Rothe, D.S. Shafer. Multiple bifurcation in a predator-prey system with non-monotonic predator response, {\it P. Roy. Soc. Edinb.} {\bf 120A}(1992) 313--347.

\bibitem{[17]}S.G. Ruan, D.M. Xiao. Global analysis in a predator-prey system with nonmonotonic function response, {\it SIAM. J. Appl. Math.} {\bf 61}(4)(2001) 1445--1472.

\bibitem{[18]}J.C. Huang, D.M. Dong. Analyses of bifurcations and stability in a predator-prey system with Holling Type-IV functional response, {\it Acta Math. Appl. Sin.-E} {\bf 20}(1)(2004) 167--178.

\bibitem{[26]}S.L. Wang, F. Xu. Threshold and bistability in HIV infection models with oxidative stress. Submitted to Journal.

\bibitem{[19]}J.M. Conway, A.S. Perelson. Post-treatment control of HIV infection, {\it Pro. Natl. Acad. Sci. USA} {\bf 112}(2015) 5467--5472.

\bibitem{[21]}H.K. Khalil. Nonlinear System, {\it Prentice-Hall} 1996.

\bibitem{[22]}J.P. La Salle. The stability of dynamical systems, {\it SIAM} 1976.

\bibitem{[23]}S. Bonhoeffer, M. Rembiszewski, G.M. Ortiz, D.F. Nixon. Risks and benefits of structured antiretroviral drug therapy interruptions in HIV-1 infection, {\it AIDS} {\bf 14}(2000) 2313--2322.



\bibitem{[25]}S.L. Wang, F. Xu. Thresholds and bistability in virus-immune dynamics, {\it Appl. Math. Lett.} {\bf 78}(2018) 105--111.


\bibitem{[27]}{J. Sotomayor. Generic bifurcation of dynamical system, {\it Dynam. Syst.} {\bf 561} (1973).}

\bibitem{[28]}{L. Perko. Differential equation and dynamical system, {\it Speinger-Verlag}, New York, {\bf 7} (2001).}

\bibitem{[29]}{M. Haque. Ratio-dependent predator-prey models of interacting populations, {\it Bull. Math. Biol.} {\bf 71} (2009)430--452.}



\end{thebibliography}

\end{document}